\documentclass[12pt,letterpaper,oneside]{article}
\usepackage[margin=1in]{geometry}
\usepackage{float}

\let\bibhang\relax
\usepackage{natbib}
\def\newblock{\hskip .11em plus.33em minus.07em}

\usepackage{graphicx,subfigure}

\usepackage{amsmath, hyperref, ulem}

\usepackage[dvipsnames]{xcolor}

\usepackage{authblk}
\title{Deep learning for low frequency extrapolation of multicomponent data in elastic full waveform inversion}
\author{Hongyu Sun and Laurent Demanet}
\affil{Massachusetts Institute of Technology, 77 Massachusetts Ave, Cambridge, MA 02139,\\ hongyus@mit.edu, laurent@math.mit.edu}
\date{December, 2020}

\begin{document}

\label{firstpage}

\maketitle

\begin{abstract}
Full waveform inversion (FWI) strongly depends on an accurate starting model to succeed. This is particularly true in the elastic regime: The cycle-skipping phenomenon is more severe in elastic FWI compared to acoustic FWI, due to the short S-wave wavelength. In this paper, we extend our work on extrapolated FWI (EFWI) by proposing to synthesize the low frequencies of multi-component elastic seismic records, and use those ``artificial" low frequencies to seed the frequency sweep of elastic FWI. Our solution involves deep learning: we separately train the same convolutional neural network (CNN) on two training datasets, one with vertical components and one with horizontal components of particle velocities, to extrapolate the low frequencies of elastic data. The architecture of this CNN is designed with a large receptive field, by either large convolutional kernels or dilated convolution. Numerical examples on the Marmousi2 model show that the 2-4Hz low frequency data extrapolated from band-limited data above 4Hz provide good starting models for elastic FWI of P-wave and S-wave velocities. Additionally, we study the generalization ability of the proposed neural network over different physical models. For elastic test data, collecting the training dataset by elastic simulation shows better extrapolation accuracy than acoustic simulation, i.e., a smaller generalization gap.


\end{abstract}


\section{Acknowledgments}
The authors thank Total SA for support. Laurent Demanet is also supported by AFOSR grant FA9550-17-1-0316. Hongyu Sun acknowledges SEG scholarships (SEG 75th Anniversary Scholarship and Michael Bahorich/SEG Scholarship) for funding. Tensorflow \citep[]{tensorflow2015-whitepaper} and Keras \citep[]{chollet2015keras} are used for deep learning. Elastic FWI in this note is implemented using the open source code DENISE (https://github.com/daniel-koehn/DENISE-Black-Edition). Acoustic training datasets are simulated using Pysit \citep{hewett2013pysit}.

\section{Introduction}

Full waveform inversion is well-known for its great potential to provide quantitative Earth properties of complex subsurface structures. Acoustic FWI is widely used and has been successfully applied to real seismic data. However, most seismic data have strong elastic effects \citep{marjanovic2018elastic}. The acoustic approximation is insufficient to estimate correct reflections and introduces additional artifacts to FWI results \citep{plessix2013multiparameter,stopin2014multiparameter}. Therefore, it is desirable to develop a robust elastic FWI method for high-resolution Earth model building. 


Foundational work has shown the ability of elastic FWI to retrieve realistic properties of the subsurface \citep{tarantola1986strategy, mora1987nonlinear}. However, it has difficulty handling real data sets. Elastic FWI is very sensitive to: accuracy of the starting model; correct estimation of density; proper definition of multi-parameter classes; and noise level \citep{brossier2010data}. The complex wave phenomena in elastic wavefields bring new challenges to FWI.

Among the many factors that affect the success of elastic FWI, the lowest starting frequency is an essential one, given that an accurate starting model is generally unavailable. Compared to acoustic FWI, the nonlinearity of elastic FWI is more severe due to the short S-wave propagating wavelength. Therefore, elastic FWI always requires a lower starting frequency compared to acoustic FWI. Additionally, the parameter cross-talk problem exists in elastic FWI and becomes more pronounced at higher frequencies, so ultra-low frequencies are required for a successful inversion of S-wave velocity and density. 

In synthetic studies of elastic FWI, \cite{brossier2009seismic} invert the overthrust model \citep{aminzadeh19973} from 1.7Hz. \cite{brossier2010data} invert the Valhall model \citep{sirgue20093d} from 2Hz. Both inversion workflows start from Gaussian smoothing of true models. Moreover, \cite{choi2008frequency} invert the Marmousi2 model \citep{martin2006marmousi2} using a velocity-gradient starting model but a very low frequency (0.16Hz). For a successful inversion of the Marmousi2 density model, \cite{kohn2012influence} use 0-2Hz in the first stage of multi-scale FWI. \cite{jeong2012full} invert the same model from 0.2Hz. 

Few applications of elastic FWI to real data sets are reported in the literature \citep{crase1990robust, sears2010elastic, marjanovic2018elastic}. \cite{vigh2014elastic} use 3.5Hz as the starting frequency of elastic FWI given that the initial models are accurate enough. \cite{raknes2015three} apply 3D elastic FWI to update P-wave velocity and obtain S-wave velocity and density using empirical relationships. \cite{borisov2020application} perform elastic FWI involving surface waves in the band of 5-15Hz for a land data set. 

New developments in acquisition enhance the recent success of FWI by measuring data with lower frequencies and longer offsets \citep{mahrooqi2012land, brenders2018wolfspar}. However, only acoustic FWI was applied to the land data set with low frequencies down to 1.5 Hz \citep{plessix2012full}. In addition to the expensive acquisition cost for the low-frequency signals, direct use of the field low-frequency data requires dedicated pre-processing steps, including travel-time tomography, for an accurate enough model to initialize FWI. The final inversion results strongly rely on the starting tomography model. Hence, attempting to retrieve reliable low-frequency data offers a sensible pathway to relieve the dependency of elastic FWI on starting models. 

Deep learning is an emerging technology in many aspects of exploration geophysics. In seismic inversion, several groups have experimented with directly mapping data to model using deep learning \citep{araya2018deep,yang2019deep,wu2019inversionnet,zhang2020data,kazei2020elastic}. Within Bayesian seismic inversion framework, deep learning has been applied for formulating priors \citep{herrmann2019learned, mosser2020stochastic, fang2020deep}. Other groups use deep learning as a signal processing step to acquire reasonable data for inversion. For instance, \cite{li2019deep} use deep learning to remove elastic artifacts for acoustic FWI. \cite{siahkoohi2019importance} remove the numerical dispersion of wavefields by transfer learning.

Computationally extrapolating the missing low frequencies from band-limited data is the cheapest way for FWI to mitigate the cycle-skipping problem. \cite{li2015phase,li2016full} seperate the shot gather to atomic events and then change the wavelet to extrapolate the low frequencies. \cite{li2017extrapolated} extend the frequency spectrum based on the redundancy of extended forward modeling. Recently, \cite{sun2018low, ovcharenko2018low, jin2018learn} have utilized CNN to extrapolate the missing low frequencies from band-limited data. They have proposed different architectures of CNN to learn the mapping between high and low frequency data from different features in the training datasets. However, only acoustic data are considered in these studies.

Although the mechanism of deep learning is hard to explain, the feasibility of low frequency extrapolation has been discussed in terms of sparsity inversion \citep{hu2019progressive} and wavenumber illumination \citep{ovcharenko2019deep}. With multiple-trace extrapolation, the low wavenumbers of far-offset data have been proposed as the features in the frequency domain detected by CNN to extrapolate the missing low frequencies \citep{ovcharenko2019deep}. In contrast, for trace-by-trace extrapolation \citep{sun2020extrapolated}, the features to learn are the structured time series themselves. The feasibility of trace-by-trace frequency extrapolation has been mathematically proved in simple settings in \cite{demanet2015recoverability, demanet2019stable}, as a by-product of super-resolution. 

In this paper, we extend our workflow of extrapolated FWI with deep learning \citep{sun2020extrapolated} into the elastic regime. We separately train the same neural network on two different training datasets, one to predict the low-frequency data of the horizontal components ($v_x$) and one to predict the low frequencies of the vertical components ($v_y$). The extrapolated low frequency data are used to initialize elastic FWI from a crude starting model. For the architecture design of CNN, a large receptive field is achieved by either large convolutional kernels or dilation convolution. Moreover, to investigate the generalization ability of neural networks over different physical models, we compare the extrapolation results of the neural networks trained on elastic data and acoustic data to predict the elastic low-frequency data. We also investigate several hyperparameters of deep learning to understand its bottleneck for low frequency extrapolation, such as mini-batch size, learning rate and the probability of dropout.

The organization of this article is as follows. In Section~\ref{section:method}, we first briefly review elastic FWI and its implementation in this paper. Then, we present the architecture of neural networks and training datasets for low frequency extrapolation of elastic data. Section~\ref{section:examples} describes the numerical results of low frequency extrapolation, extrapolated elastic FWI and investigation of hyperparameters. Section~\ref{section:limitation} discusses the limitations of the method. Section~\ref{section:conclusion} comes to conclusions and future directions.

\section{Method}
\label{section:method}

We first give a brief review of elastic FWI as implemented in this paper. Then we illustrate the feasibility of low frequency extrapolation, and design two deep learning models for this purpose. Afterwards, the training and test datasets are provided to train and verify the performance of the proposed neural networks. 

\subsection{Review of elastic FWI}

Elastic FWI is implemented in the time domain to invert the P-wave velocities ($\mathbf{v}_p$), S-wave velocities ($\mathbf{v}_s$) and density ($\rho$) simultaneously. The object function $E$ is formulated as 
\begin{equation}
E=\frac{1}{2}\delta \mathbf{d}^T \delta \mathbf{d}=\frac{1}{2}\sum\limits_s \sum\limits_r \int [\mathbf{u}_{cal}-\mathbf{u}_{obs}]^2 dt,
\label{object}
\end{equation}
where $\mathbf{d}$ are the residuals between observed wavefields $\mathbf{u}_{obs}$ and calculated wavefields $\mathbf{u}_{cal}$. In 2D, both $\mathbf{u}_{obs}$ and $\mathbf{u}_{cal}$ contain the $v_x$ and $v_y$ components of elastic wavefields. The gradient $\frac{\delta{E}}{\delta{\mathbf{m}}}$ relative to the model parameters $\mathbf{m}$ is calculated in terms of $\mathbf{v}_p$, $\mathbf{v}_s$ and $\rho$ using the velocity-stress formulation of the elastic wave equation \cite[]{kohn2012influence}. The starting models $\mathbf{m_0}$ are updated using the L-BFGS method \cite[]{nocedal2006numerical}.

\subsection{Deep learning models for low-frequency extrapolation}

We choose CNN to perform the task of low-frequency extrapolation. By trace-by-trace extrapolation, the output and input are the same seismic recording in the low and high frequency band, respectively. In 2D, the elastic data contain horizontal and vertical components. As a result, we propose to separately train the same neural network twice on two different training datasets: one contains $v_x$ and the other contains $v_y$.

We design two kinds of CNN architectures with large receptive field for low-frequency extrapolation. A very large receptive field enables each feature in the final output to include a large range of input pixels. Since any single frequency component is related to the entire waveform in the time domain, extrapolation from one frequency band to the other requires a large receptive field to cover the entire input signal. Typically, the receptive field is increased by stacking layers. For example, stacking 
two convolutional layers (without pooling) with $3\times3$ filter results in a layer of $5\times5$ filter. However, it requires many layers to result in a large enough receptive field and is computationally inefficient. Therefore, we design the CNN architecture with two methods: directly use a large filter or a small filter with dilated convolution.

\begin{figure}
 \includegraphics[width=\columnwidth]{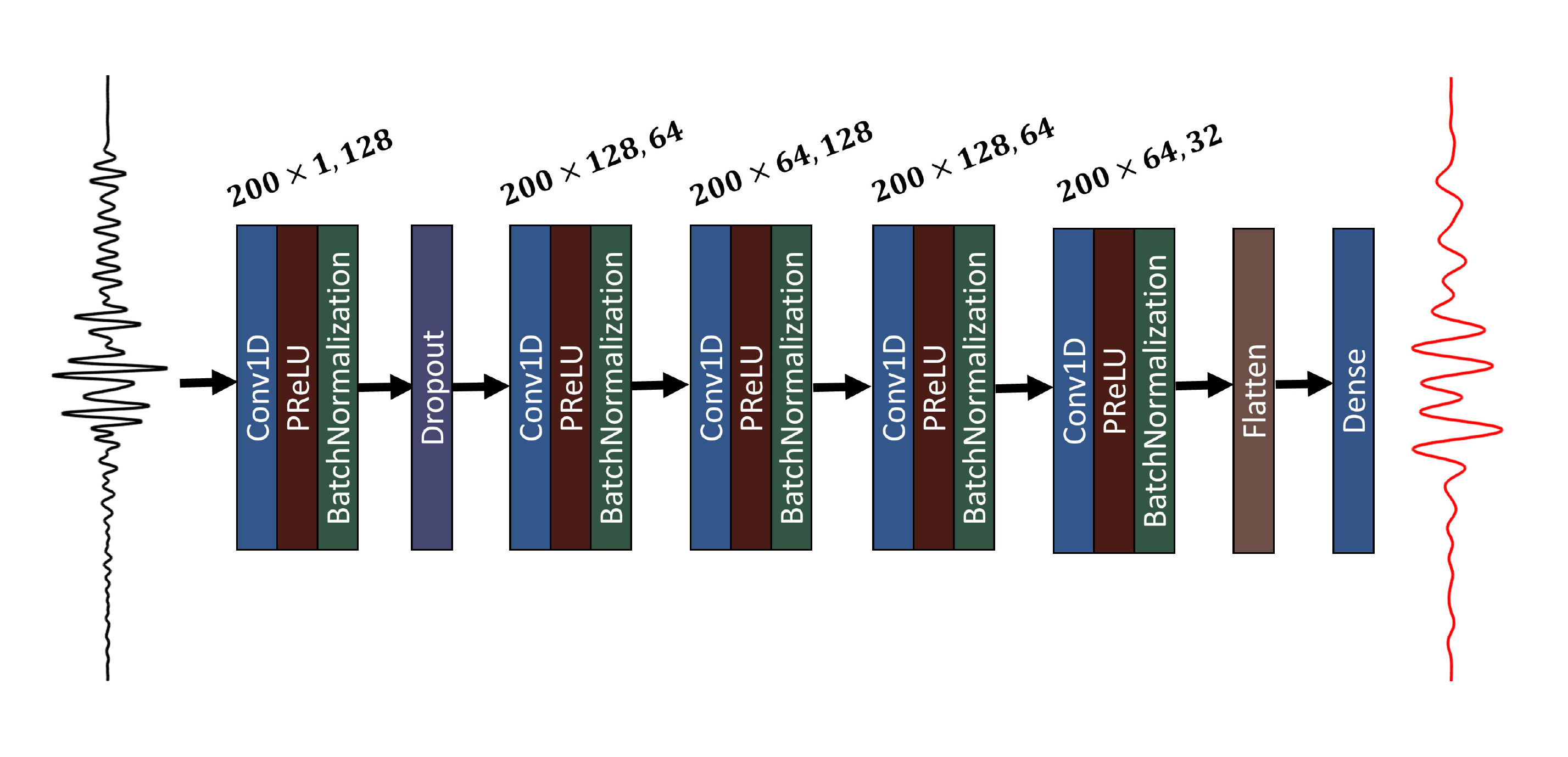}
   \caption{ARCH1. The first choice of deep learning architecture that directly employs a large filter on each convolutional layer \citep{sun2020extrapolated}. The size and number of filters are labeled on the top of each convolutional layer.}
\label{fig:arch1}
\end{figure}

The first CNN architecture (ARCH1) directly employs a large filter on convolutional layers, which is the same as \citet{sun2020extrapolated} (Figure~\ref{fig:arch1}). Recall that ARCH1 is a feed-forward stack of five convolutional blocks. Each block is a combination of a 1D convolutional layer, a PReLU layer and a batch normalization layer. The length of all filters is 200. On each convolutional layer, the channels number 64, 128, 64, 128 and 32, respectively. Although only one trace is plotted in Figure~\ref{fig:arch1}, the proposed neural network can easily explore multiple traces of the shot gather for multi-trace extrapolation by increasing the size of the kernels from $200\times1$ to $200\times ntr$, where $ntr$ is the number of the input traces. 

\begin{figure}
  \centering
  \includegraphics[width=0.7\textwidth]{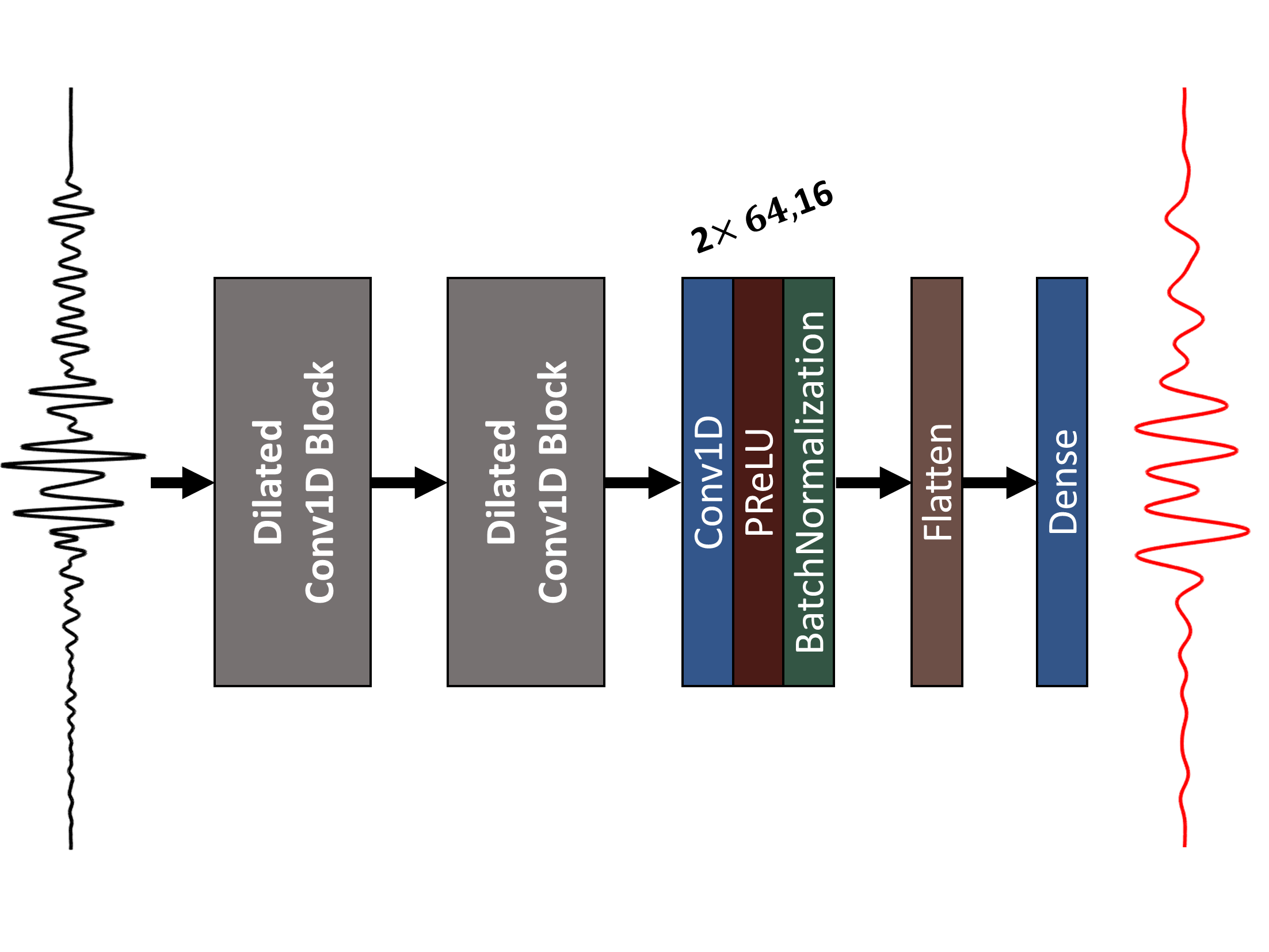} \\
  \includegraphics[width=0.8\textwidth]{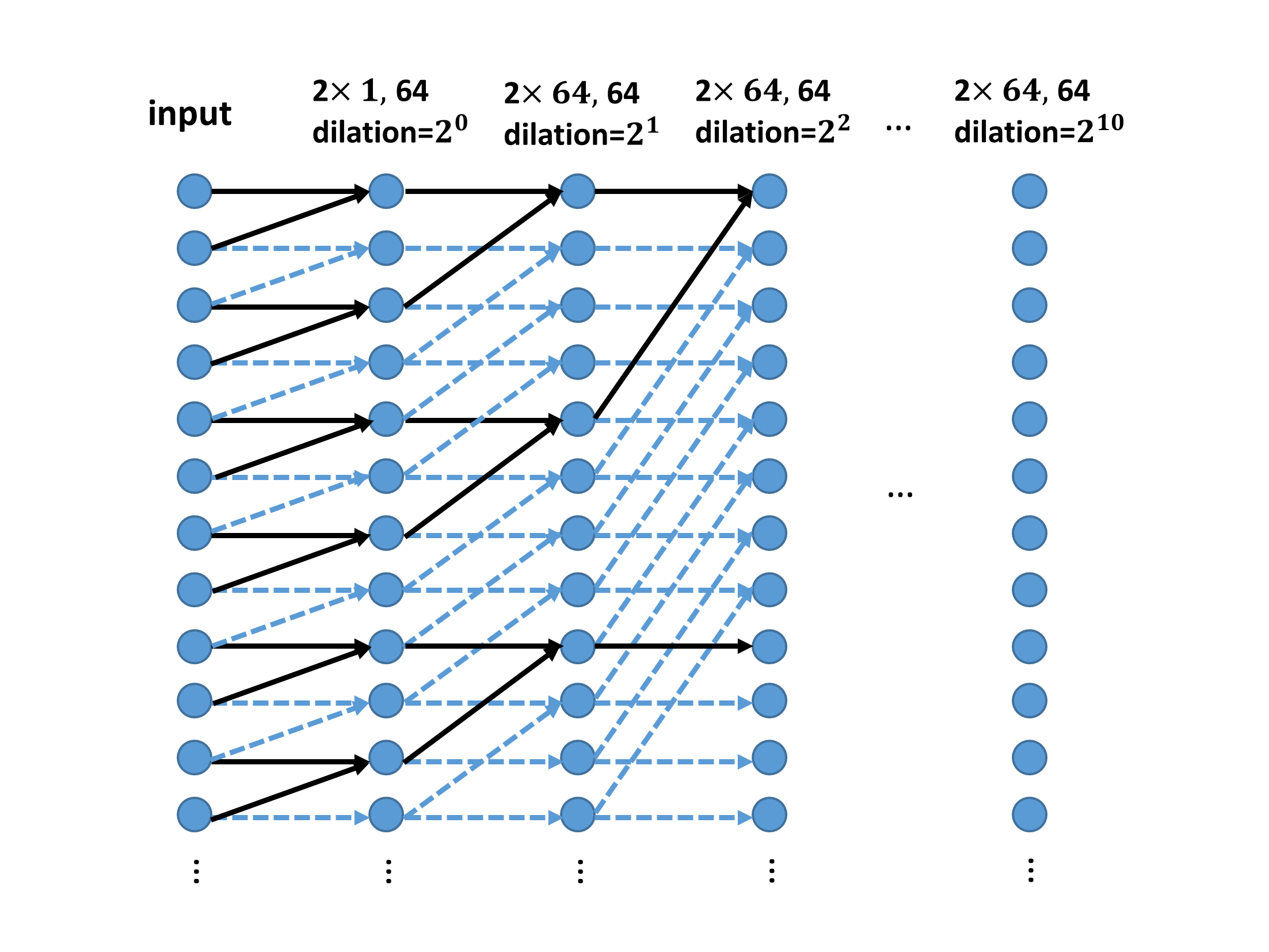}
  \caption{(a)ARCH2. The second choice of deep learning architecture with dilated convolution. The filter length is 2 on each convolutional layer but two dilated convolutional blocks are stacked to increase the receptive field exponentially with depth. (b)The dilated causal convolutional block. Each block has 11 convolutional layers. The filter size, number of channel and dilation of each convolutional layer are labeled on the top.}
  \label{fig:arch2}
\end{figure}

The second CNN architecture (ARCH2) uses dilated convolution to increase the receptive field by orders of magnitude. A dilated convolution (convolution with holes) is a convolution where the filter is applied over an area larger than its length by skipping input values with a certain step (dilation). It effectively allows the network to operate on a coarser scale than with a normal convolution. This is similar to pooling or stride, but here the output has the same size as the input. As a special case, dilated convolution with dilation of one yields the standard convolution. Stacked dilated convolutions enable networks to have very large receptive fields with just a few layers. In addition to save computational cost, this method helps to preserve the input resolution throughout the network \citep{oord2016wavenet}. Moreover, we use causal convolution to process time series \citep{moseley2018fast}, although this choice does not appear to be essential in our case. 


The architecture of ARCH2 (Figure~\ref{fig:arch2}) has two dilated convolutional blocks. Each block consists of 10 1D convolutional layers. Each convolutional layer is followed by a PReLU layer and a batch normalization layer. On each convolutional layer, there are 64 causal convolutional filters with a length of 2 (Figure~\ref{fig:arch2}b). The dilations of the ten convolutional layers are $2^0$, $2^1$, ..., $2^{10}$, respectively. The exponential increase in dilation results in exponential growth, with depth, of the receptive field \citep{yu2015multi}. 

With the two proposed architectures, we can compare the specific receptive fields of both ARCH1 and ARCH2. Without pooling layer, the size of the receptive field ${RF}_{l+1}$ on the $l+1$ layer is 
\begin{equation}
\label{eq:receptive field}
{RF}_{l+1}={RF}_{l}+(k_{l+1}-1)\times s_{l+1} \times d_{l+1}, \quad l=0,...,n \quad ,
\end{equation}
\begin{equation}
{RF}_0=1,
\end{equation}
where $k_{l+1}$ is the kernel size of the $l+1$ convolutional layer. $s_{l+1}$ is the stride size. $d_{l+1}$ is the dilation on the $l+1$ layer if the layer contains a dilated convolution. Otherwise, $d_{l+1}$ equals 1 for regular convolutional layers.

The receptive field is 996 for ARCH1 and 4095 for ARCH2. Since a smaller kernel is used in ARCH2, the number of trainable parameter in ARCH2 ($p_2$=148,453,480) is much less than ARCH1 ($p_1$=294,602,648). Although both neural networks are able to perform low frequency extrapolation, convolution with dilation is more efficient than directly using a large convolutional kernel. 

\subsection{Training and test datasets}
\label{section:Training and test datasets}

\begin{figure}
 \includegraphics[width=\columnwidth]{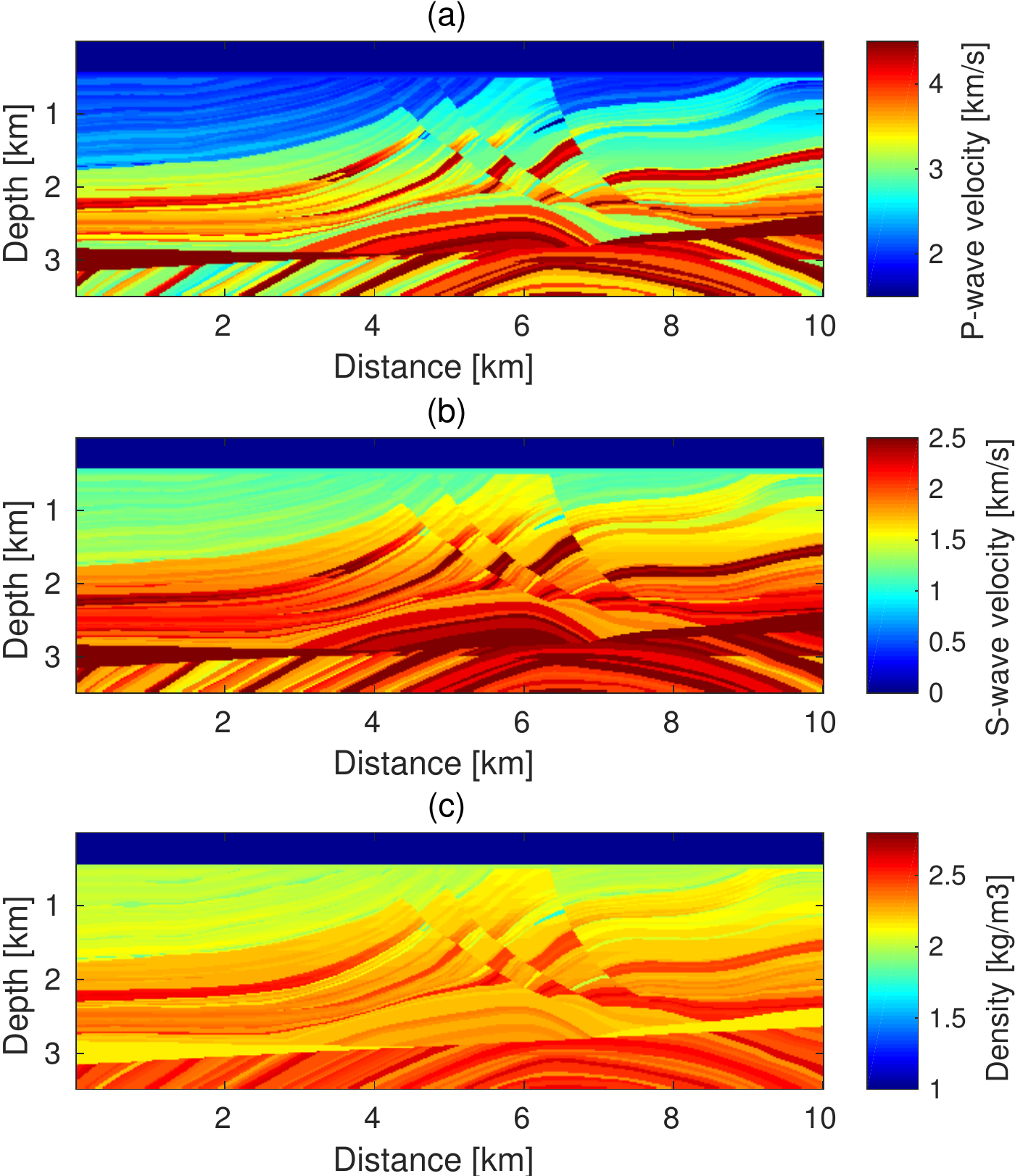}
   \caption{The true Marmousi2 model: (a)$\mathbf{v}_p$, (b)$\mathbf{v}_s$ and (c)$\rho$.}
   \label{fig:marmousi_II_marine}
\end{figure}

\begin{figure}
 \includegraphics[width=\columnwidth]{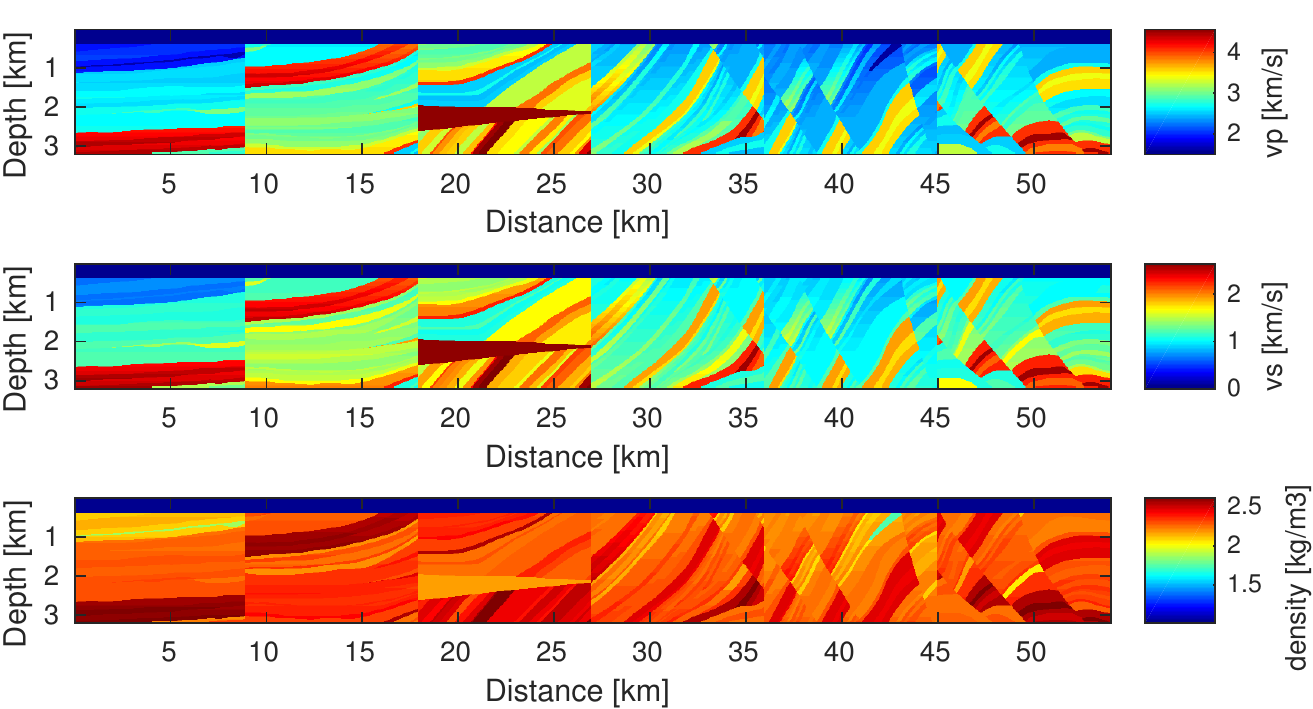}
   \caption{The six training models randomly extracted from the Marmousi2 model. The size of each training model is equally $500\times174$ with a grid spacing of 20m including a 440m depth water layer. Each training model contains three parameters: $\mathbf{v}_p$, $\mathbf{v}_s$ and $\mathbf{\rho}$.}
   \label{fig:training_elastic_model}
\end{figure}

The training and test datasets are simulated on the elastic training and test models. The Marmousi2 elastic model (Figure~\ref{fig:marmousi_II_marine}) is referred to as the test model in deep learning. This is also the true model in the subsequent elastic FWI. The training models (Figure~\ref{fig:training_elastic_model}) are six batches randomly extracted from the Marmousi2 model. Our previous work
 \citep[]{sun2020extrapolated} has shown that the random selection of the training models on the Marmousi model provides enough generalization ability for the neural network to extrapolate low frequencies on the full-size Marmousi model. Here in the elastic regime, each model consists of three parameters: $\mathbf{v}_p$, $\mathbf{v}_s$ and $\mathbf{\rho}$. The size of each model is $500\times174$ with a grid spacing of 20m, including a water layer on the top of each model with a depth of 440m. 

Both training and test datasets are simulated using a 2D time domain stress-velocity P–SV finite-difference (FD) code \citep{virieux1986p, levander1988fourth} with an eighth-order spatial FD operator. A Ricker wavelet with a dominant frequency of 10Hz is used as the source signal. The sampling rate and the recording time is 0.02s and 6s, respectively. It is not necessary to collect the training and test datasets using the same acquisition geometry. To collect the test dataset, 50 shots are excited evenly from 800m to 8640m in the water layer at the same depth of 40m. 400 receivers are placed from 800m to 8780m under the water layer with a depth of 460m to record $v_x$ and $v_y$ of the elastic wavefields. However, for the training model, there are 100 shots excited evenly from 500m to 8420m with a depth of 40m on each training model. 400 receivers are placed from 480m to 8460m. 

After the forward modeling, two training datasets are collected, one with a dataset of horizontal components and one with a dataset of vertical components. The 2D elastic data on the test model is also separated into two test datasets to process each component individually. By trace-by-trace extrapolation setup, there are $6\times100\times400=240,000$ training samples in each training dataset and $1\times50\times400=20,000$ test samples in each test dataset. 

A simple preprocessing step can be used to improve the deep learning performance. Each sample in the training and test datasets is normalized to one by dividing the raw signal by its maximum. Then all the data are scaled with a constant (for instance, 100) to stabilize the training process. The values used to normalize and scale the raw data are recorded to recover the original observed data for elastic FWI. After this process, each sample in the training and test dataset is separated into a low-frequency signal and a high-frequency signal using a smooth window in the frequency domain. Then, each time series in the high-frequency band is fed into the neural network to predict the low-frequency time series. 


\section{Numerical Examples}
\label{section:examples}

The numerical examples section is divided into four parts. In the first part, we train ARCH1 to extrapolate the low-frequency data of bandlimited multi-component recordings simulated on the Marmousi2 model (Figure~\ref{fig:marmousi_II_marine}). In the second part, we study the generalization ability of the proposed neural network over different physical models (acoustic or elastic wave equation). Then, we use the extrapolated low-frequencies of multi-component band-limited data to seed the frequency sweep of elastic FWI on the Marmousi2 model. In the last part, we investigate the hyperparameters of deep learning for low-frequency extrapolation using the proposed ARCH1 and ARCH2.

\subsection{Low frequency extrapolation of multicomponent data}

\begin{figure}
\centering     
\subfigure[Figure A]{\label{fig:vy_training_loss}\includegraphics[width=60mm]{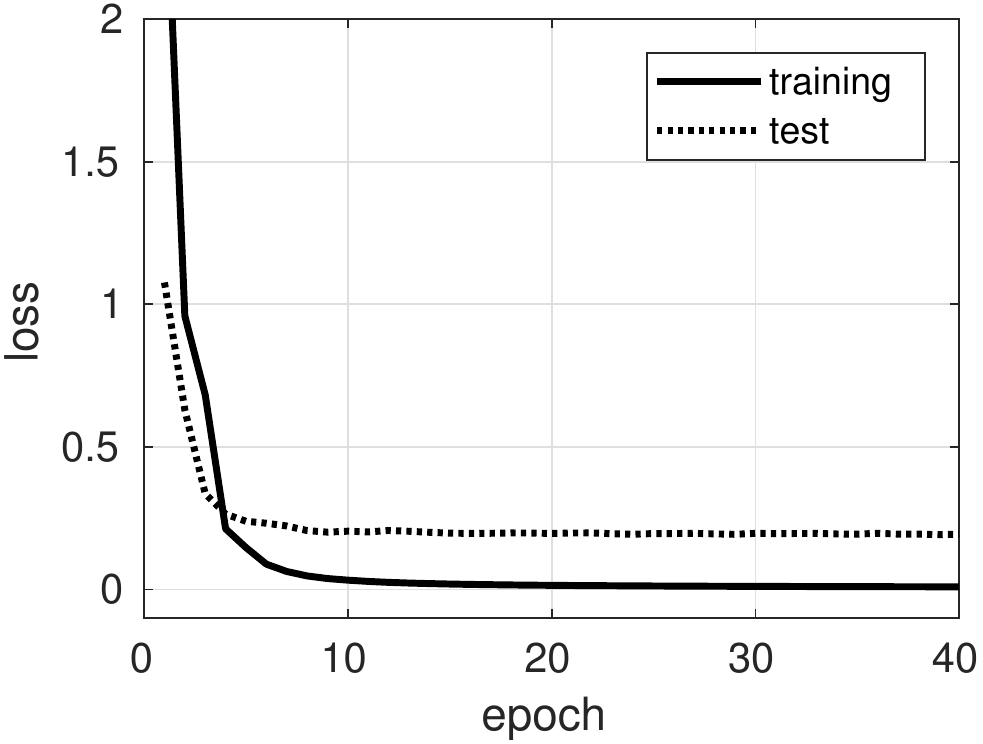}}
\subfigure[Figure B]{\label{fig:vx_training_loss}\includegraphics[width=60mm]{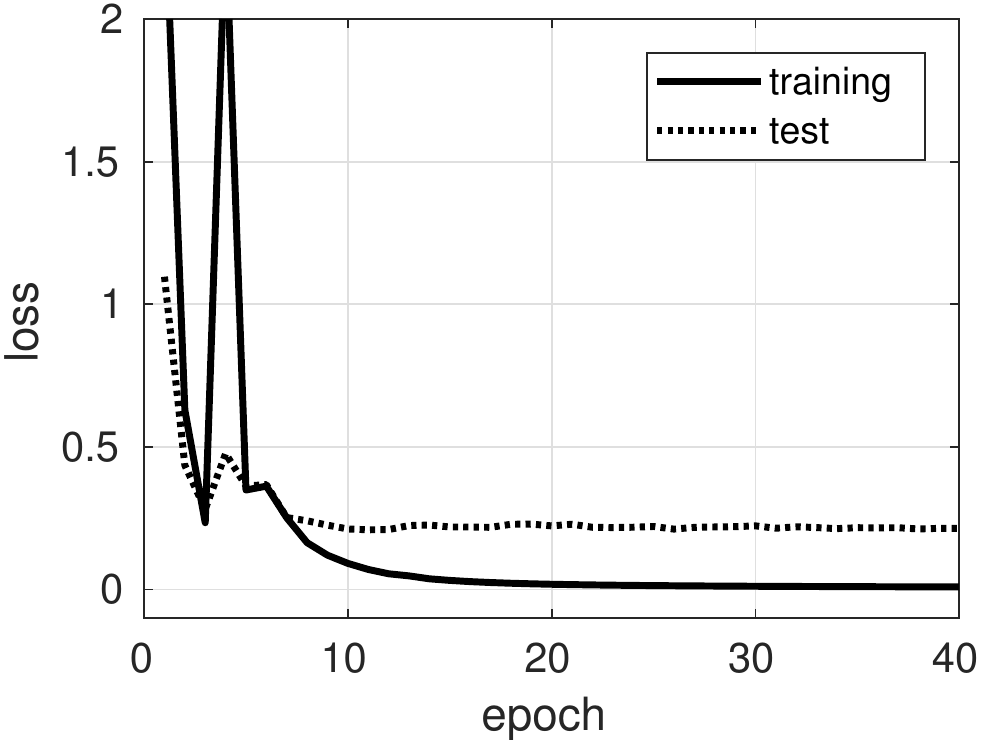}}
\caption{The learning curves of ARCH1 trained to extrapolate the 0-5Hz low frequencies of (a) $v_y$ and (b) $v_x$ from the 5-25Hz band-limited elastic recordings.}
\label{fig:training_loss}
\end{figure}

We first extrapolate the low frequency data below 5Hz on the Marmousi2 model (Figure~\ref{fig:marmousi_II_marine}) using 5-25Hz band-limited data. Each sample in the training and test datasets is separated into a 0-5Hz low-frequency signal and a 5-25Hz high-frequency signal using a smooth window in the frequency domain. The time series in the high-frequency band is directly fed into the neural network to predict the 0-5Hz low-frequency time series. To deal with the multicomponent data, the neural network ARCH1 is trained twice: once on the training dataset of $v_x$ and once on the training dataset of $v_y$. Both training processes use the ADAM method with a mini-batch of 32 samples. We refer readers to \citet[]{sun2020extrapolated} for more details about training. Figures~\ref{fig:vy_training_loss} and~\ref{fig:vx_training_loss} show the training processes over 40 epochs to predict the low frequencies of $v_x$ and $v_y$, respectively. The curves of training loss decay over epochs on both the training and test datasets, which indicate that the neural network does not overfit.

\begin{figure*}
 \includegraphics[width=\columnwidth]{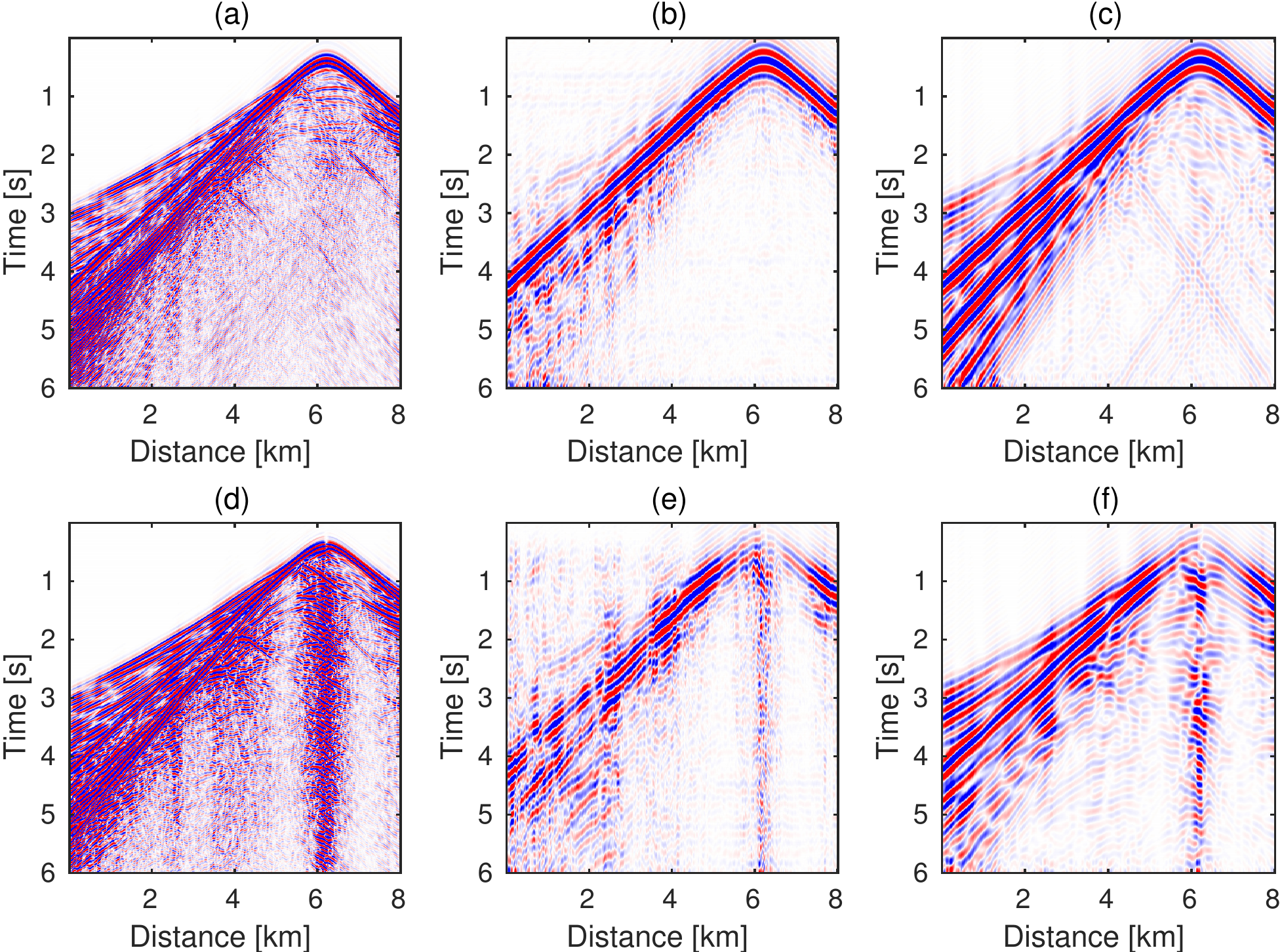}
   \caption{The extrapolation results of ARCH1 for Marmousi2 model: comparison among the (a) band-limited recordings ($5.0-25.0$Hz), (b) predicted and (c) true low-frequency recordings ($0.1-5.0$Hz) of $v_y$ and (d) band-limited recordings ($5.0-25.0$Hz), (e) predicted and (f) true low-frequency recordings ($0.1-5.0$Hz) of $v_x$.}
   \label{fig:records_ishot_40_elastic_all}
\end{figure*}

\begin{figure}
\centering    
\subfigure[Figure A]{\label{fig:spectrum_line_vy_ishot_40_itrace_266}\includegraphics[width=60mm]{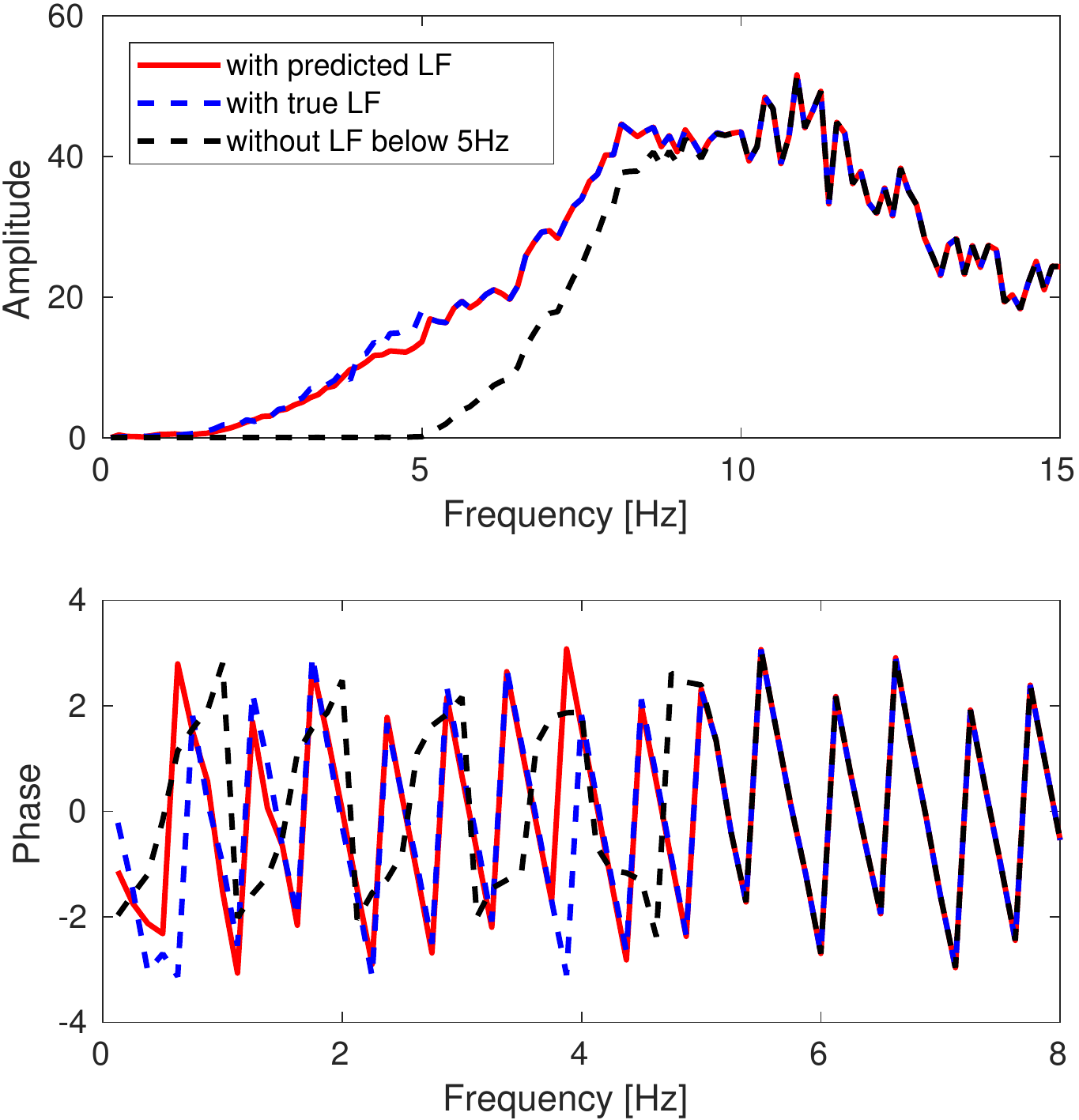}}
\subfigure[Figure B]{\label{fig:spectrum_line_vx_ishot_40_itrace_266}\includegraphics[width=60mm]{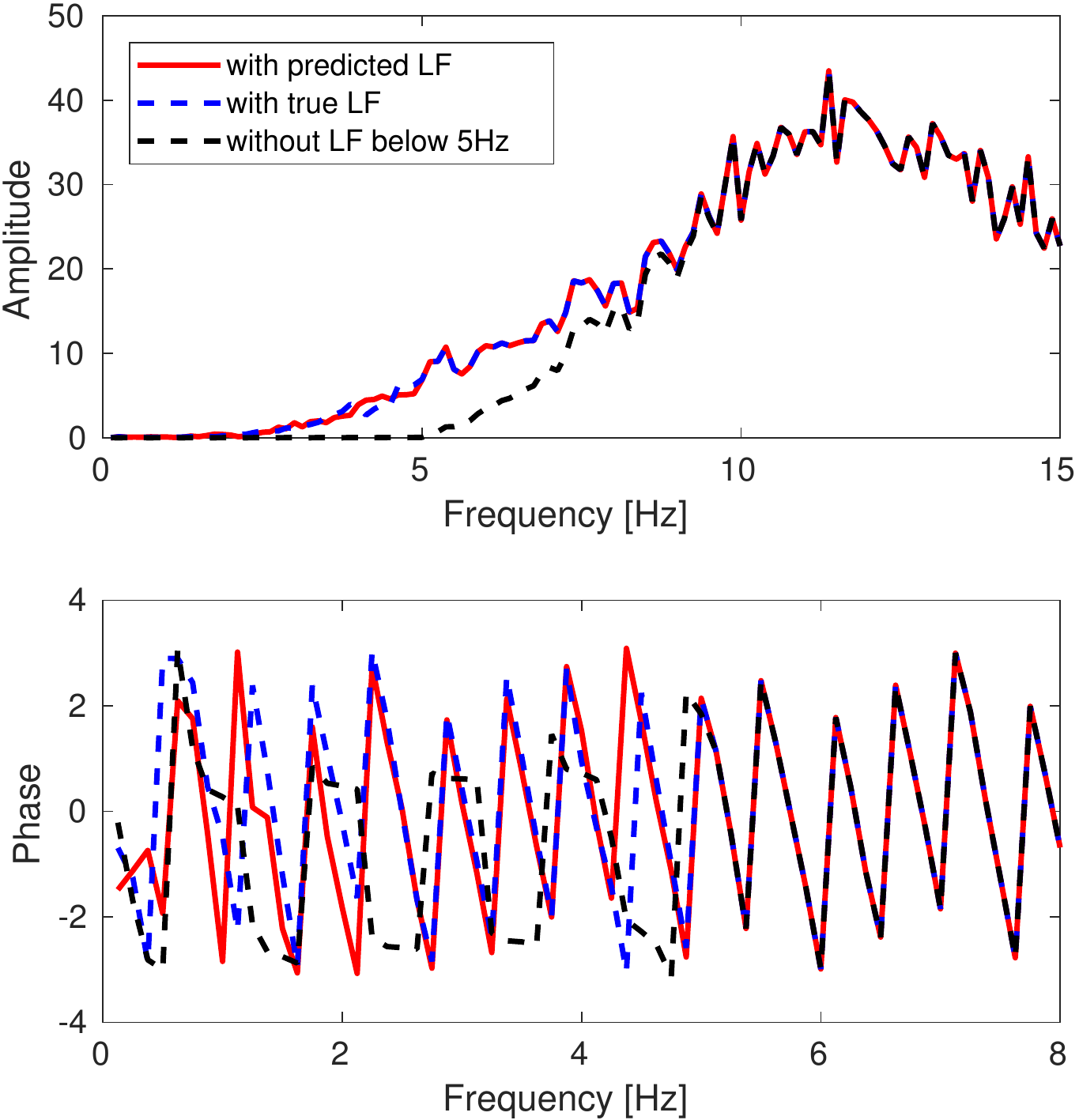}}
\caption{Extrapolation results of ARCH1 \textit{trained on elastic data}: comparison of the amplitude and phase spectrum of (a) $v_y$ and (b) $v_x$ at $x=6.1$km among the bandlimited recording ($5.0-25.0$Hz), the recording ($0.1-25.0$Hz) with true and predicted low frequencies ($0.1-5.0$Hz).}
\label{fig:elastic_spectrum}
\end{figure}


Figure~\ref{fig:records_ishot_40_elastic_all} shows the extrapolation results of both $v_x$ and $v_y$ where the source is located at 7.04km. Figures~\ref{fig:spectrum_line_vy_ishot_40_itrace_266} and~\ref{fig:spectrum_line_vx_ishot_40_itrace_266} compare the amplitude and phase spectrum of $v_y$ and $v_x$ at $x=6.1$km among the band-limited recording ($5.0-25.0$Hz), the fullband recording with true and predicted low frequencies ($0.1-5.0$Hz). Despite minor prediction errors on both amplitude and phase, the neural network ARCH1 can successfully recover the low frequencies of $v_x$ and $v_y$ recordings with satisfactory accuracy.

\begin{figure}
\centering    
\subfigure[Figure A]{\label{fig:acoustic_spectrum_line_vy_ishot_40_itrace_266}\includegraphics[width=60mm]{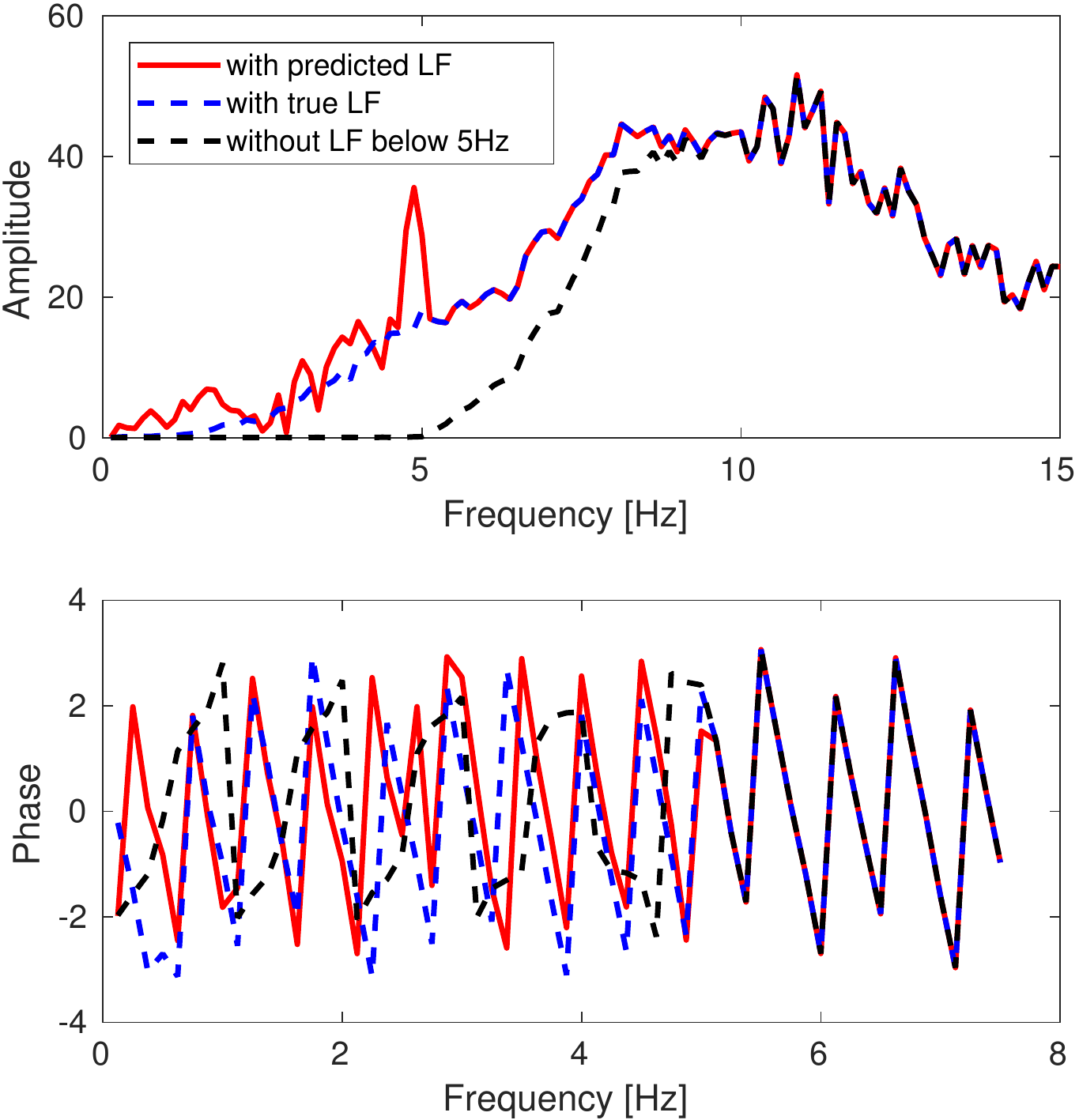}}
\subfigure[Figure B]{\label{fig:acoustic_spectrum_line_vx_ishot_40_itrace_266}\includegraphics[width=60mm]{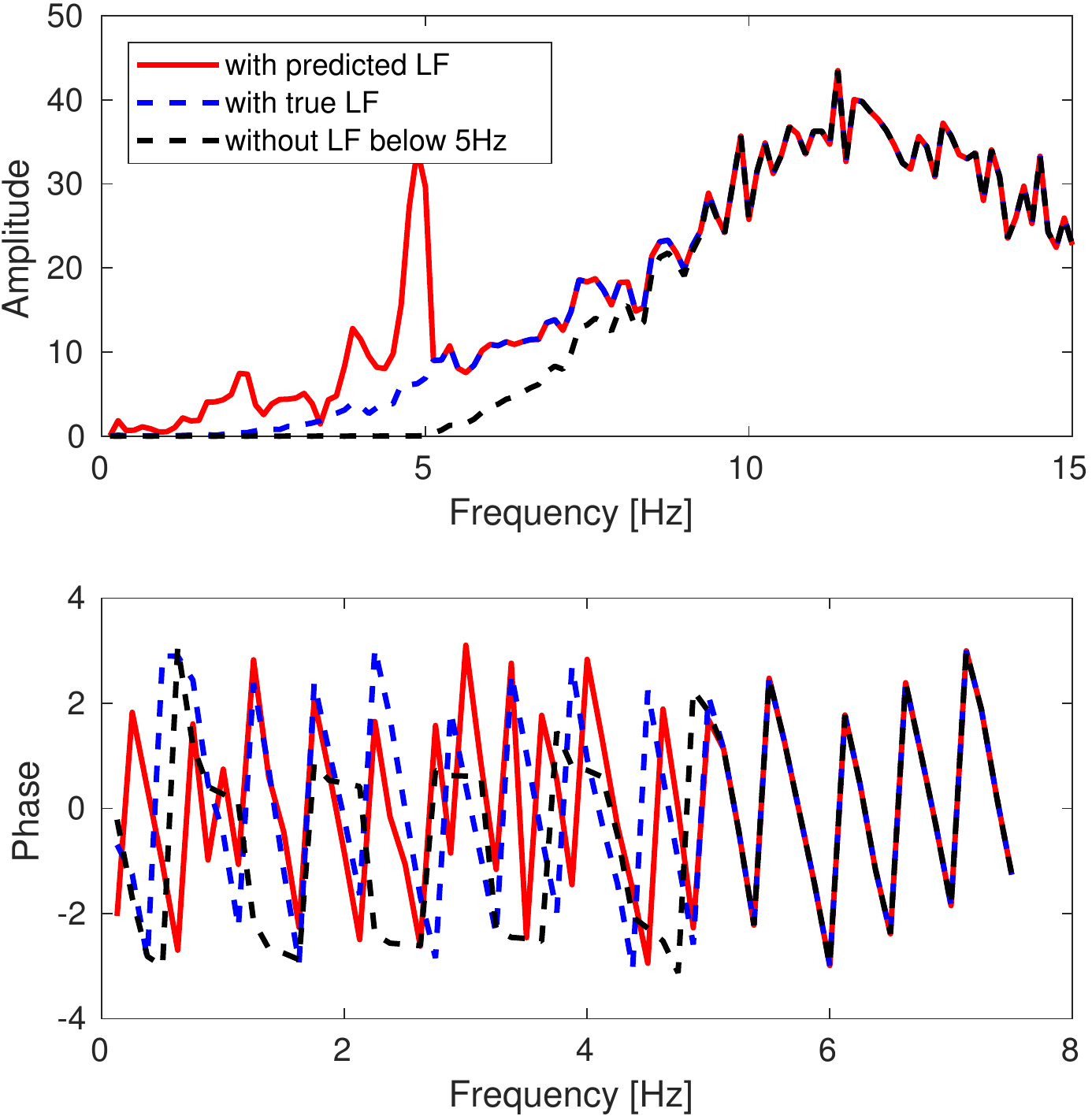}}
\caption{Extrapolation results of ARCH1 \textit{trained on acoustic data}: comparison of the amplitude and phase spectrum of (a) $v_y$ and (b) $v_x$ at $x=6.1$km among the bandlimited recording ($5.0-25.0$Hz), the recording ($0.1-25.0$Hz) with true and predicted low frequencies ($0.1-5.0$Hz).}
\label{fig:acoustic_spectrum}
\end{figure}

\subsection{Generalization over physical models}
To study the generalization ability of the proposed neural network over different physical models (acoustic wave and elastic wave), we train ARCH1 once on an acoustic training dataset and simultaneously predict the low frequencies of both $v_x$ and $v_y$ in the same elastic test dataset. The acoustic training dataset is simulated using the acoustic wave equation on only the P-wave velocity models in Figure~\ref{fig:training_elastic_model}. Figures~\ref{fig:acoustic_spectrum_line_vy_ishot_40_itrace_266} and~\ref{fig:acoustic_spectrum_line_vx_ishot_40_itrace_266} show the amplitude and phase spectrum of $v_y$ and $v_x$ at $x=6.1$km after training with the same procedure. Compared with the results in Figure~\ref{fig:elastic_spectrum}, the extrapolation accuracy of the same trace in the test data is much poorer on acoustic training dataset than elastic training dataset. Even the extrapolation of the vertical component is not successful when the neural network is trained on acoustic data. This is an indicator that the neural network has difficulty generalizing to different physical models.

\subsection{Extrapolated elastic full waveform inversion}
\label{section:extrapolated elastic FWI}

\begin{figure}
 \includegraphics[width=\columnwidth]{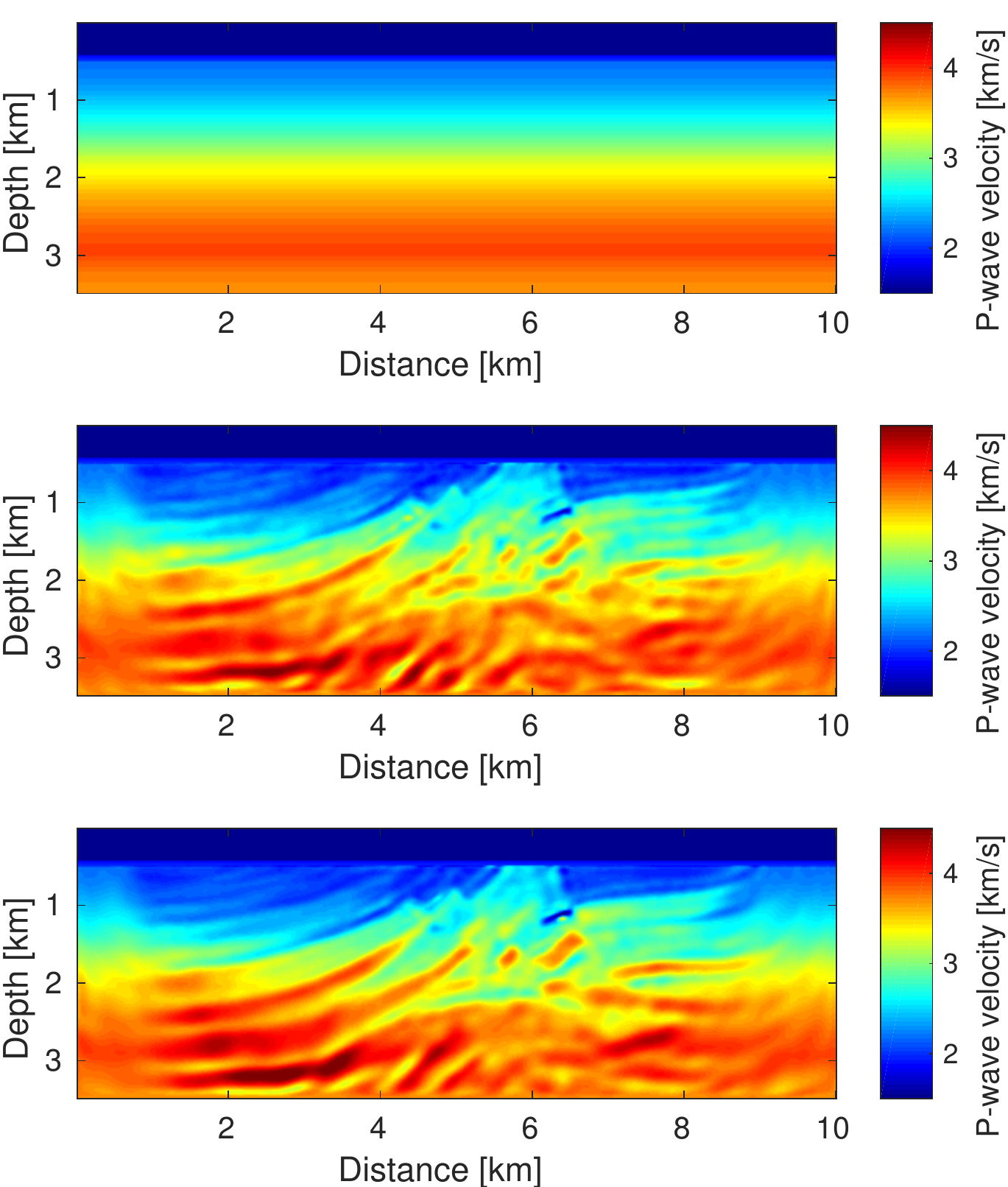}
   \caption{Comparison among (a) the initial $\mathbf{v}_p$ model, the inverted low-wavenumber velocity models using (b) $2.0-4.0Hz$ extrapolated data and (c) $2.0-4.0Hz$ true data. The inversion results in (b) and (c) are started from the initial model in (a).}
   \label{fig:vp_inversion_result_stage_1_it_30}
\end{figure}

\begin{figure}
 \includegraphics[width=\columnwidth]{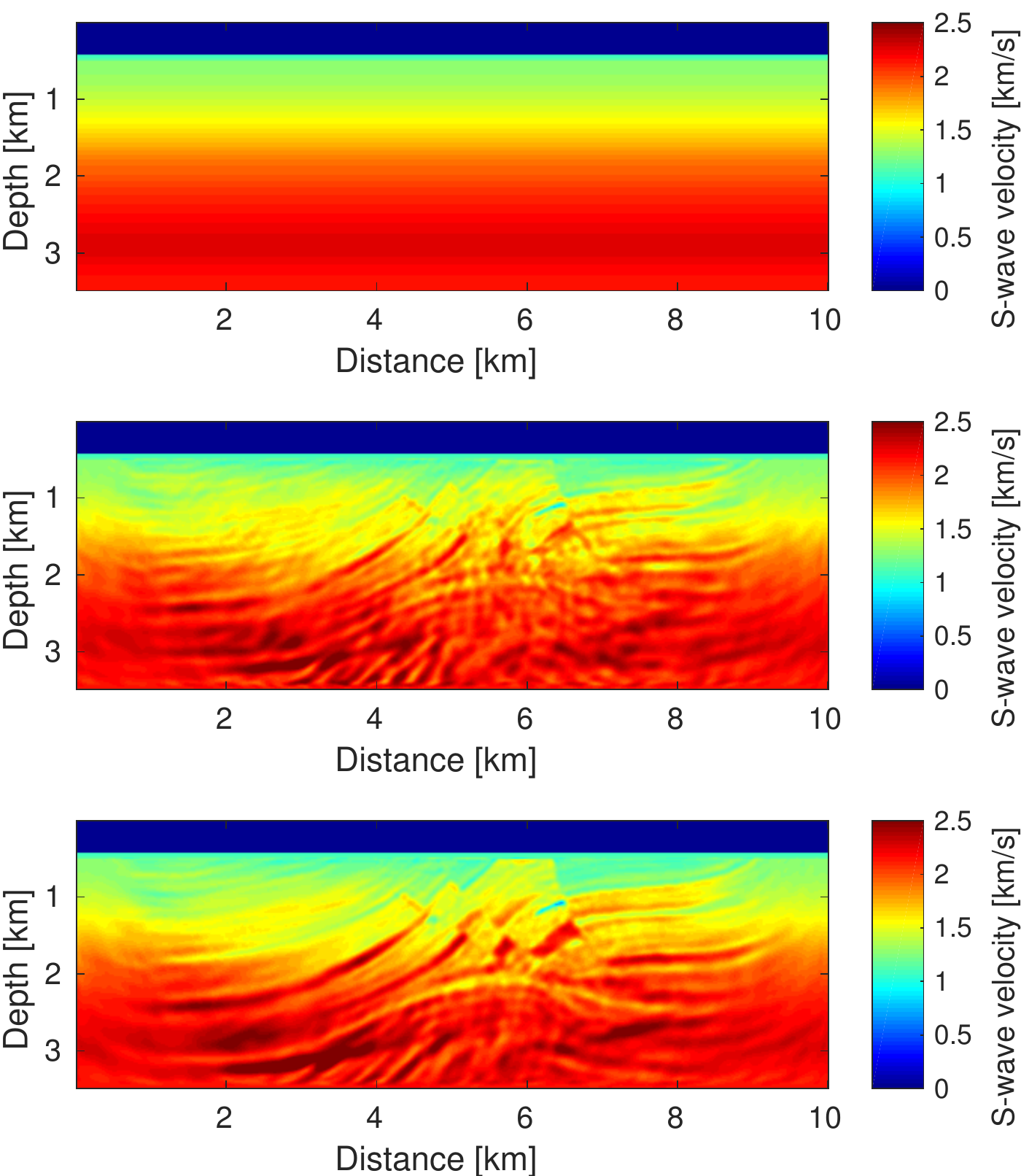}
   \caption{Comparison among (a) the initial $\mathbf{v}_s$ model, the inverted low-wavenumber velocity models using (b) $2.0-4.0Hz$ extrapolated data and (c) $2.0-4.0Hz$ true data. The inversion results in (b) and (c) are started from the initial model in (a).}
   \label{fig:vs_inversion_result_stage_1_it_30}
\end{figure}

\begin{figure}
 \includegraphics[width=\columnwidth]{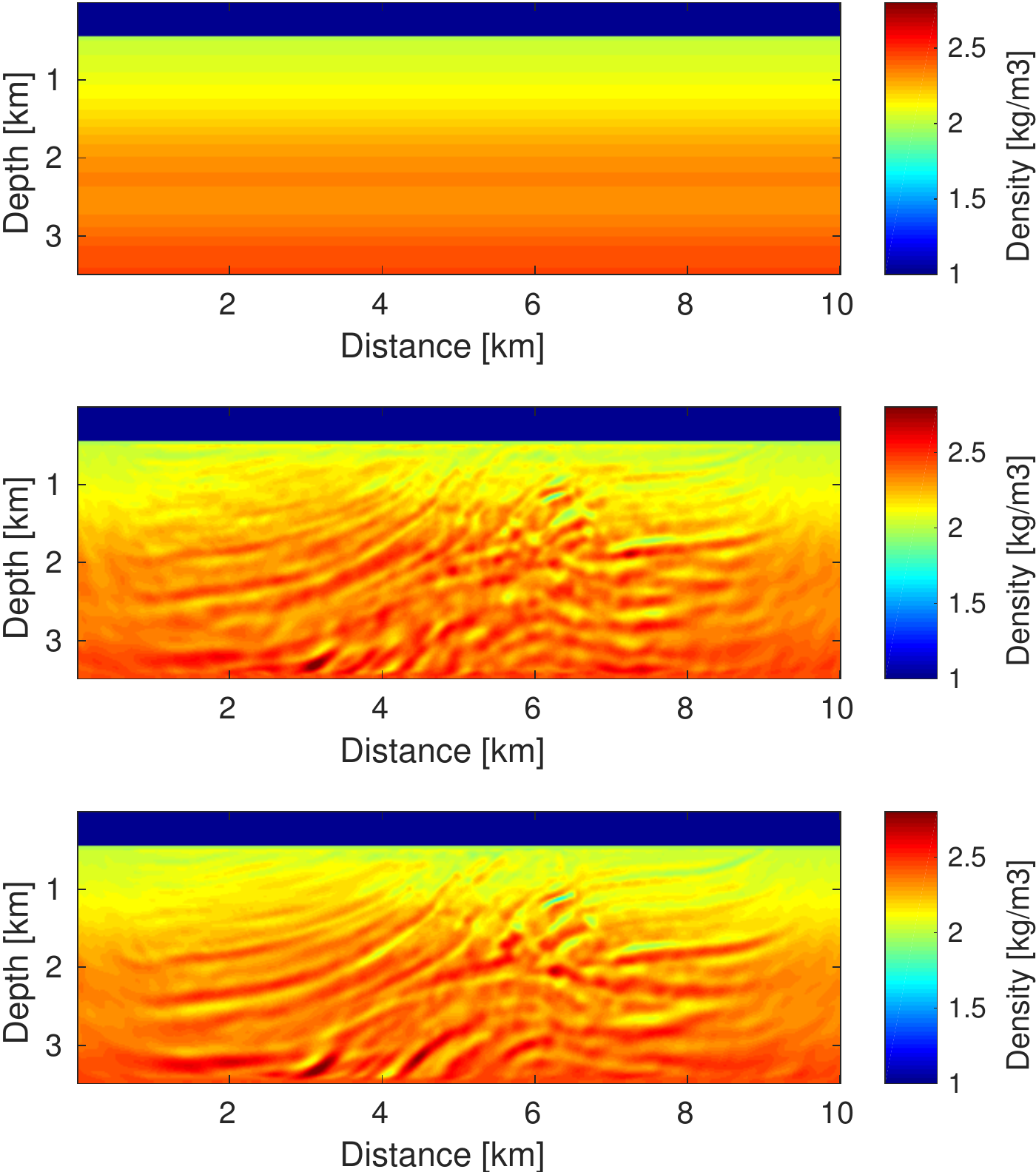}
   \caption{Comparison among (a) the initial $\rho$ model, the inverted low-wavenumber density models using (b) $2.0-4.0Hz$ extrapolated data and (c) $2.0-4.0Hz$ true data. The inversion results in (b) and (c) are started from the initial model in (a).}
   \label{fig:rho_inversion_result_stage_1_it_30}
\end{figure}

\begin{figure}
 \includegraphics[width=\columnwidth]{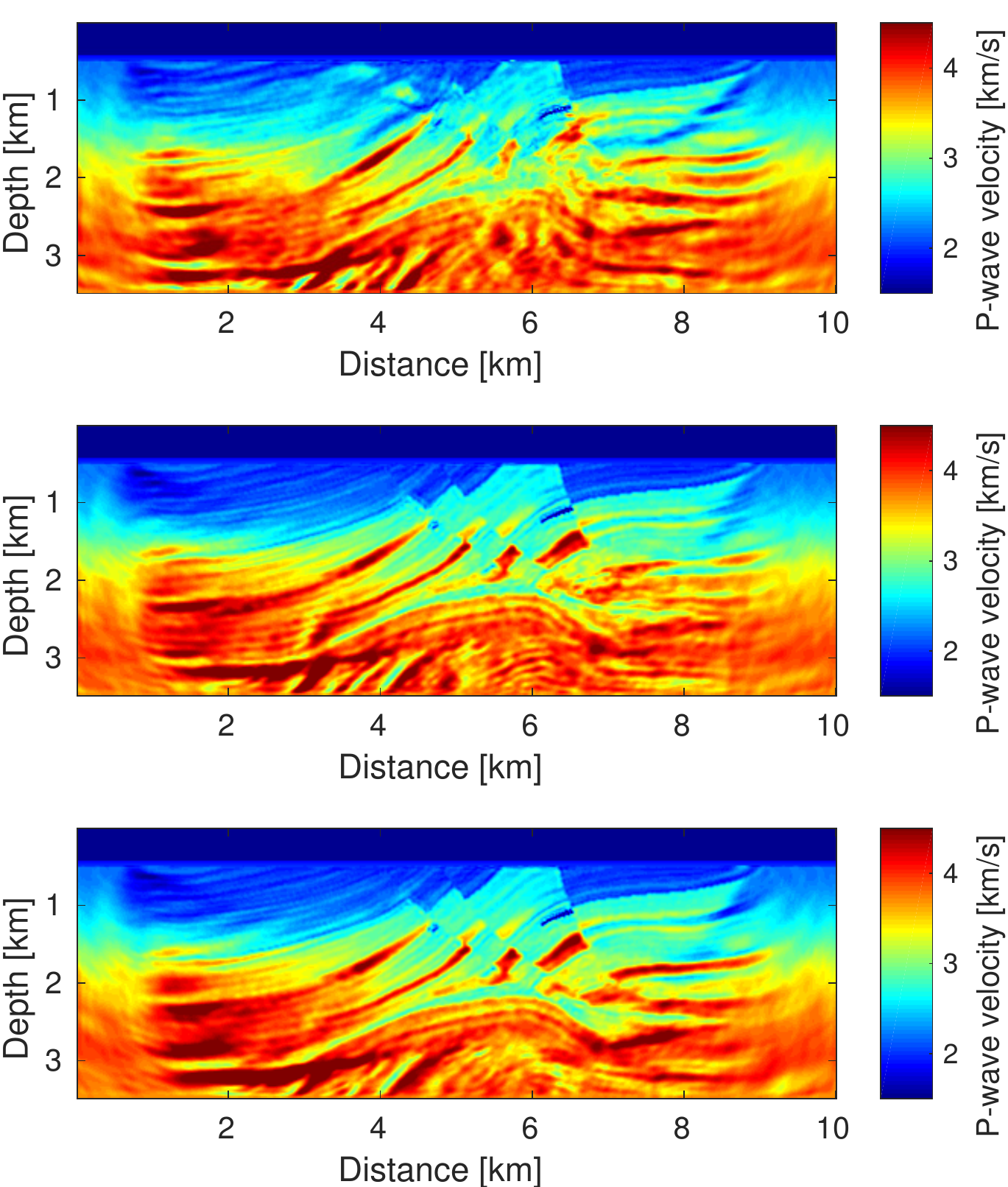}
   \caption{Comparison of the inverted $\mathbf{v}_p$ models from elastic FWI using $4-20Hz$ band-limited data. (a) The resulting model starts from the original initial model. (b) The resulting model starts from the inverted low-wavenumber velocity model using $2.0-4.0Hz$ extrapolated data. (c) The resulting model starts from the inverted low-wavenumber velocity model using $2.0-4.0Hz$ true data.}
   \label{fig:vp_inversion_result_stage_4_it_20}
\end{figure}

\begin{figure}
 \includegraphics[width=\columnwidth]{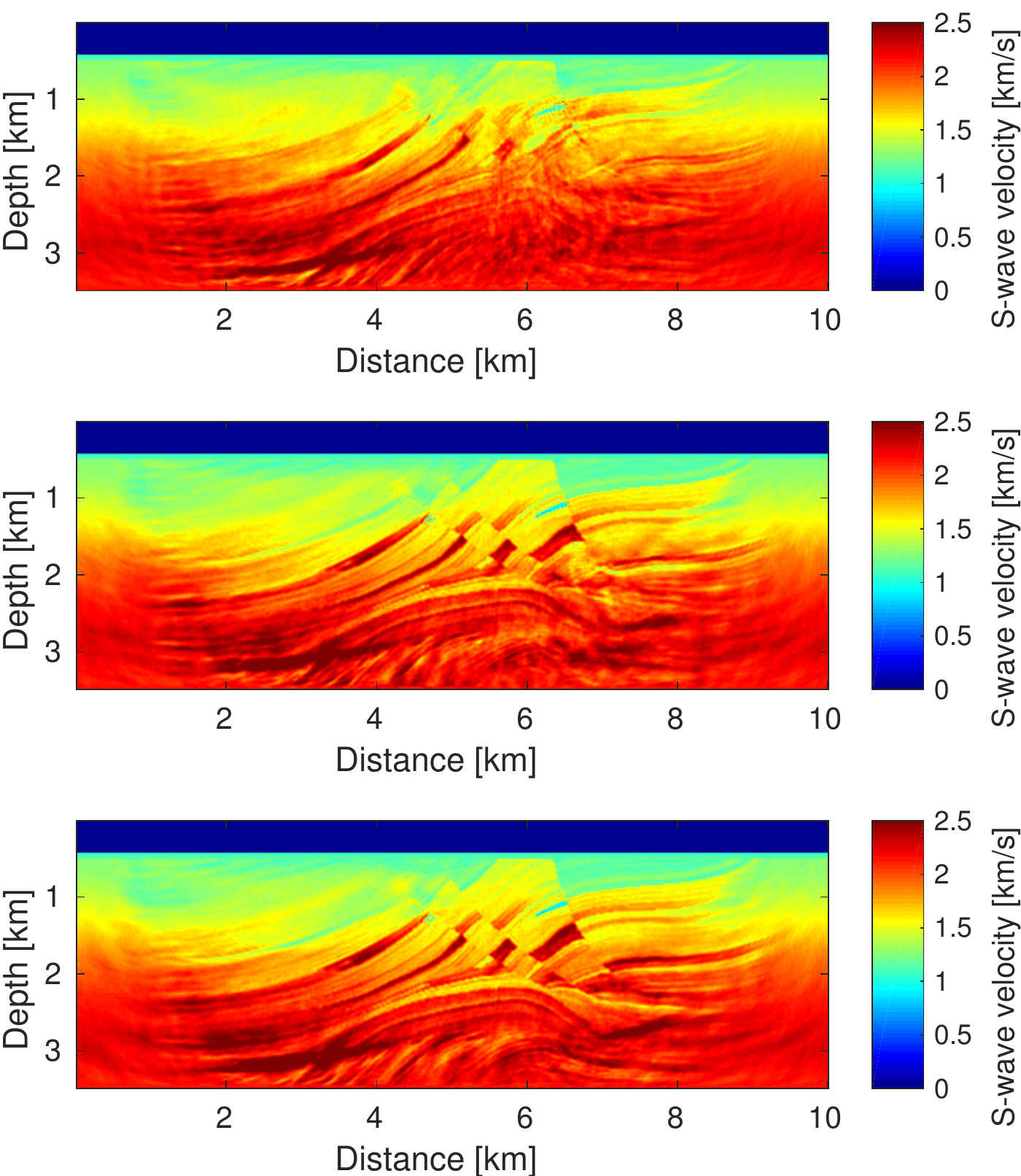}
   \caption{Comparison of the inverted $\mathbf{v}_s$ models from elastic FWI using $4-20Hz$ band-limited data. (a) The resulting model starts from the original initial model. (b) The resulting model starts from the inverted low-wavenumber velocity model using $2.0-4.0Hz$ extrapolated data. (c) The resulting model starts from the inverted low-wavenumber velocity model using $2.0-4.0Hz$ true data.}
   \label{fig:vs_inversion_result_stage_4_it_20}
\end{figure}

\begin{figure}
 \includegraphics[width=\columnwidth]{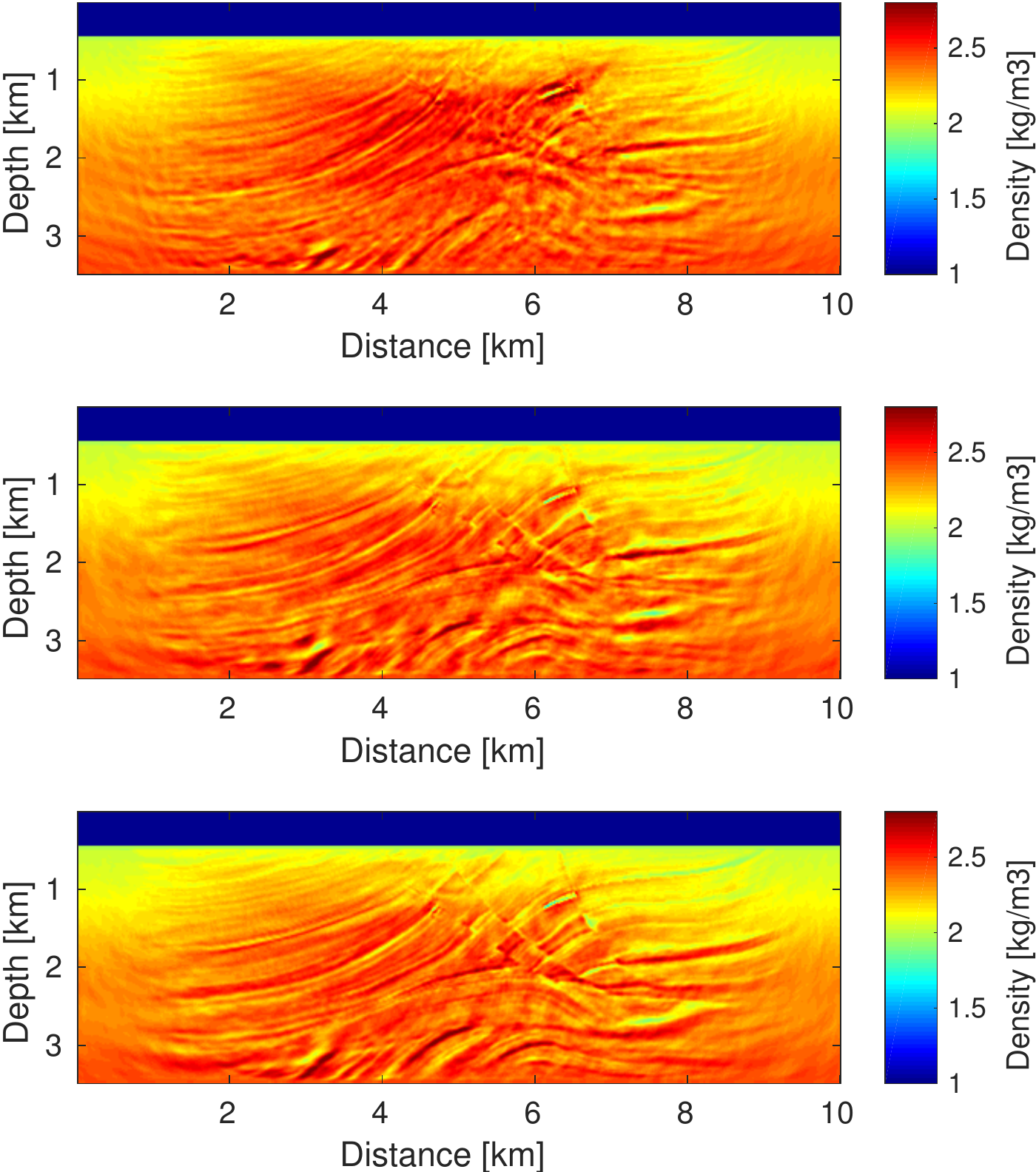}
   \caption{Comparison of the inverted $\rho$ models from elastic FWI using $4-20Hz$ band-limited data. (a) The resulting model starts from the original initial model. (b) The resulting model starts from the inverted low-wavenumber density model using $2.0-4.0Hz$ extrapolated data. (c) The resulting model starts from the inverted low-wavenumber density model using $2.0-4.0Hz$ true data.}
   \label{fig:rho_inversion_result_stage_4_it_20}
\end{figure}

We perform extrapolated elastic FWI using 4-20Hz band-limited data on the Marmousi2 model. The lower band of the band-limited data is 4Hz. Figures~\ref{fig:vp_inversion_result_stage_1_it_30}a, \ref{fig:vs_inversion_result_stage_1_it_30}a and \ref{fig:rho_inversion_result_stage_1_it_30}a show the initial models of $v_p$, $v_s$ and $\rho$, respectively. Unlike in the previous examples, a free surface boundary condition is applied to the top of the model to simulate the realistic marine exploration environment. The free surface condition damages the low frequency data and thus the energy in the low frequency band 0-2Hz is close to zero in the simulated full-band data. This brings a new challenge to the low frequency extrapolation and introduces prediction errors to the extrapolated data. For this reason, we start elastic FWI using 2-4Hz extrapolated data before exploring the band-limited data.  

Starting from the crude initial model in Figure~\ref{fig:vp_inversion_result_stage_1_it_30}(a), Figures~\ref{fig:vp_inversion_result_stage_1_it_30}(b) and \ref{fig:vp_inversion_result_stage_1_it_30}(c) show the resulting P-wave velocity models after 30 iterations using extrapolated and true 2-4Hz low-frequency data, respectively. The inverted S-wave velocity models using extrapolated and true low frequencies are shown in Figures~\ref{fig:vs_inversion_result_stage_1_it_30}(b) and \ref{fig:vs_inversion_result_stage_1_it_30}(c). Also, Figures~\ref{fig:rho_inversion_result_stage_1_it_30}(b) and \ref{fig:rho_inversion_result_stage_1_it_30}(c) compare the inverted density models using the extrapolated and true 2-4Hz low frequencies. The inverted low-wavenumber models of $v_p$, $v_s$ and $\rho$ using extrapolated data are roughly the same as those using true data. However, the inversion of density model is not successful since 2-4Hz data are relatively high frequencies for the inversion of density model.

Then the inversion is continued with the 4-20Hz band-limited data. We utilize a multiscale method \citep{bunks1995multiscale} and sequentially explore the 4-6Hz, 4-10Hz and 4-20Hz band-limited data in the elastic FWI. In each frequency band, the number of iterations is 30, 30 and 20, respectively. Figures~\ref{fig:vp_inversion_result_stage_4_it_20} and \ref{fig:vs_inversion_result_stage_4_it_20} show the resulting $v_p$ and $v_s$ models started from different low wavenumber models. The inversion results of $v_p$ and $v_s$ started from 2-4Hz extrapolated data are very close to the results started from 2-4Hz true data. Conversely, elastic FWI directly starting from the crude initial models using the band-limited data shows large errors.

Figure \ref{fig:rho_inversion_result_stage_4_it_20} shows the resulting density models using the 4-20Hz band-limited data but started from different models in Figure \ref{fig:rho_inversion_result_stage_1_it_30}. Since the starting frequency band (2-4Hz) is relatively high for the inversion of density on the crude initial model (Figure~\ref{fig:rho_inversion_result_stage_1_it_30}(a)), the inverted models using 2-4Hz data (Figures~\ref{fig:rho_inversion_result_stage_1_it_30}(b) and (c)) only show high-wavenumber structure of the density model. With band-limited data involved in the inversion, the inverted density models (Figure~\ref{fig:rho_inversion_result_stage_4_it_20}) resemble migration results but show the density perturbation \citep{mora1987nonlinear}. For a successful inversion of the density model, a much lower starting frequency band is required to recover the low-wavenumber structures.

\subsection{Investigation of hyperparameters}
\label{section:benchmark}

The performance of deep learning is very sensitive to the hyperparameters of training. However, choosing appropriate hyperparameters requires expertise and extensive trial and error. Here we discuss the influence of several hyperparameters on the training process, including mini-batch size, learning rate, and a layer-specific hyperparameter, i.e., dropout. We also compare the performance of ARCH1 and ARCH2 in terms of training cost and prediction accuracy. In each case, we compare the model performance of the new hyperparameter setting with the results predicted by the neural network ARCH1 in Section~\ref{section:extrapolated elastic FWI}. The neural network is trained using a mini-batch of 32, learning rate of $10^{-3}$. The probability of dropout is $50\%$. 

\begin{figure}
\centering     
\subfigure[]{\label{fig:benchmark_batch_vy_training_loss}\includegraphics[width=150mm]{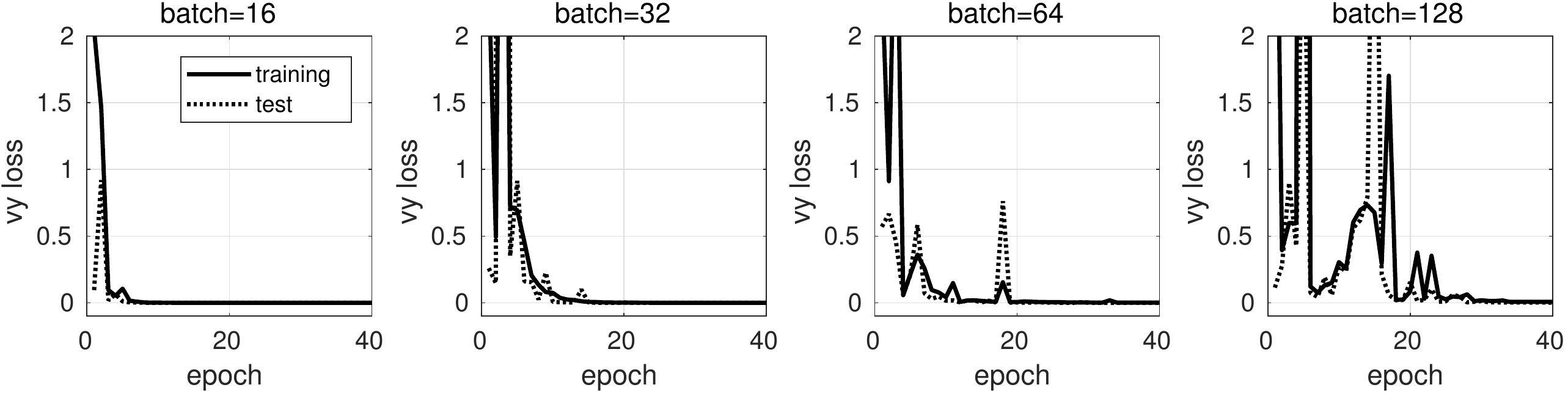}}
\subfigure[]{\label{fig:benchmark_batch_vx_training_loss}\includegraphics[width=150mm]{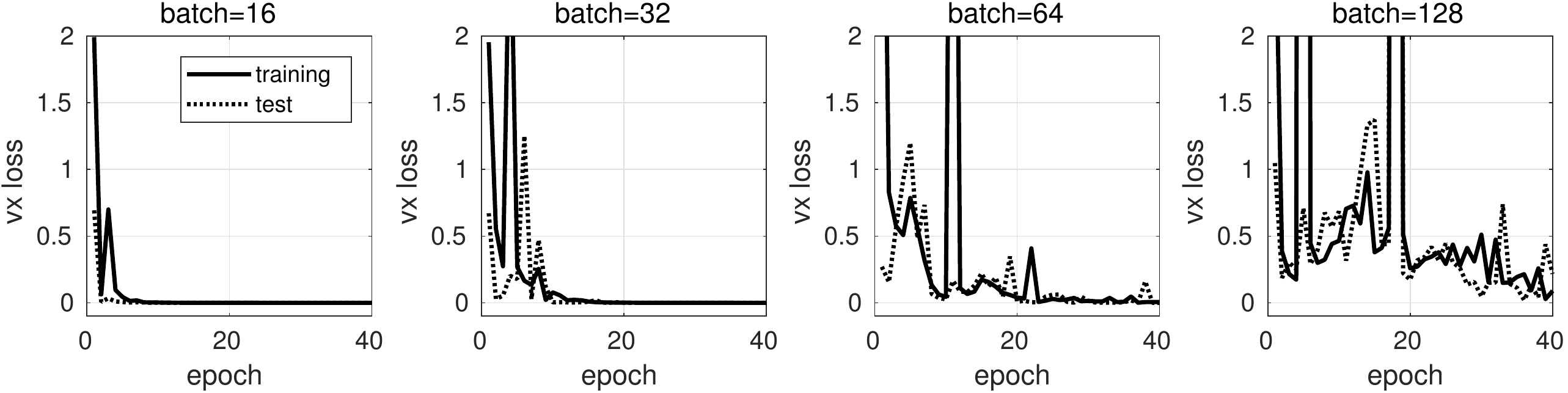}}
\caption{The learning curves of ARCH1 trained using different batch sizes to extrapolate the 0-4Hz low frequencies of (a) $v_y$ and (b) $v_x$ from the 4-25Hz band-limited elastic recordings.}
\label{fig:benchmark_batch}
\end{figure}

The first hyperparameter is the mini-batch size. A batch is a small subset of training data randomly selected by the optimizer to calculate the gradient. Choosing a suitable mini-batch size is a trade-off between training speed and test accuracy. Fewer samples in a batch slow down the training process but a large batch increases the instability of the neural network and lead to a poor performance. Figures~\ref{fig:benchmark_batch_vy_training_loss} and \ref{fig:benchmark_batch_vx_training_loss} show the learning curves of ARCH1 when processing the $v_y$ and $v_x$ components using a batch size of 16, 32, 64 and 128. A mini-batch of 32 gives more reasonable decrease of the training loss among others. Therefore, we choose it as the mini-batch size in our experiments. 

\begin{figure}
\centering    
\subfigure[]{\label{fig:benchmark_learningrate_vy_training_loss}\includegraphics[width=120mm]{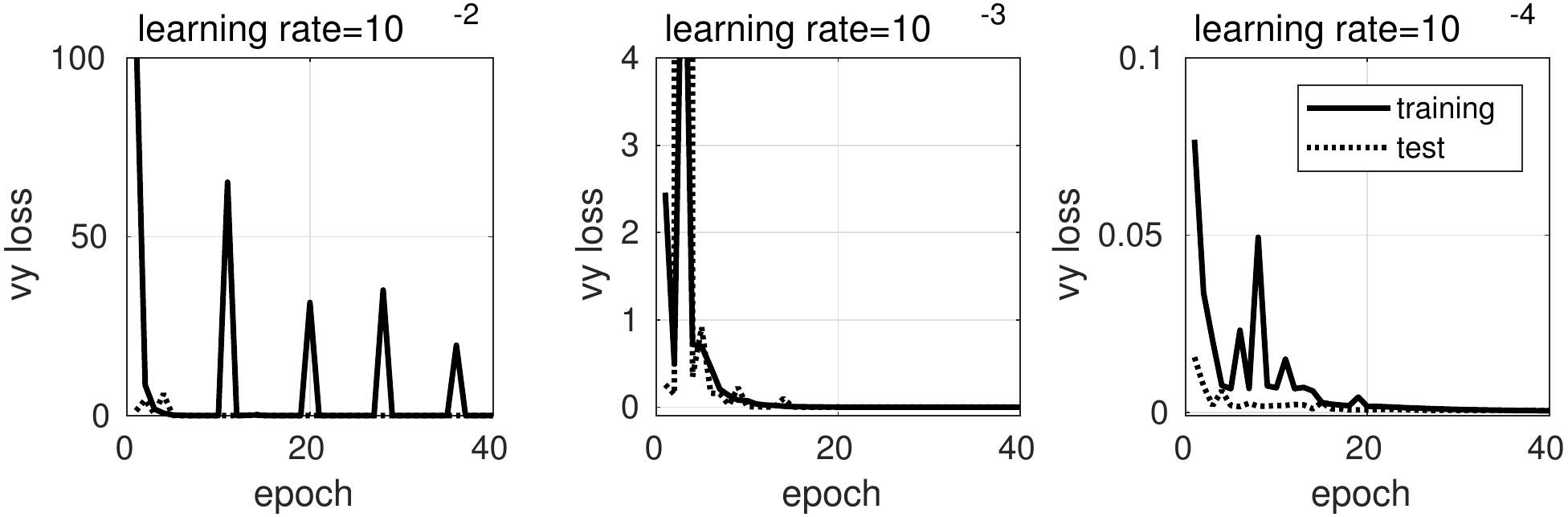}}
\subfigure[]{\label{fig:benchmark_learningrate_vx_training_loss}\includegraphics[width=120mm]{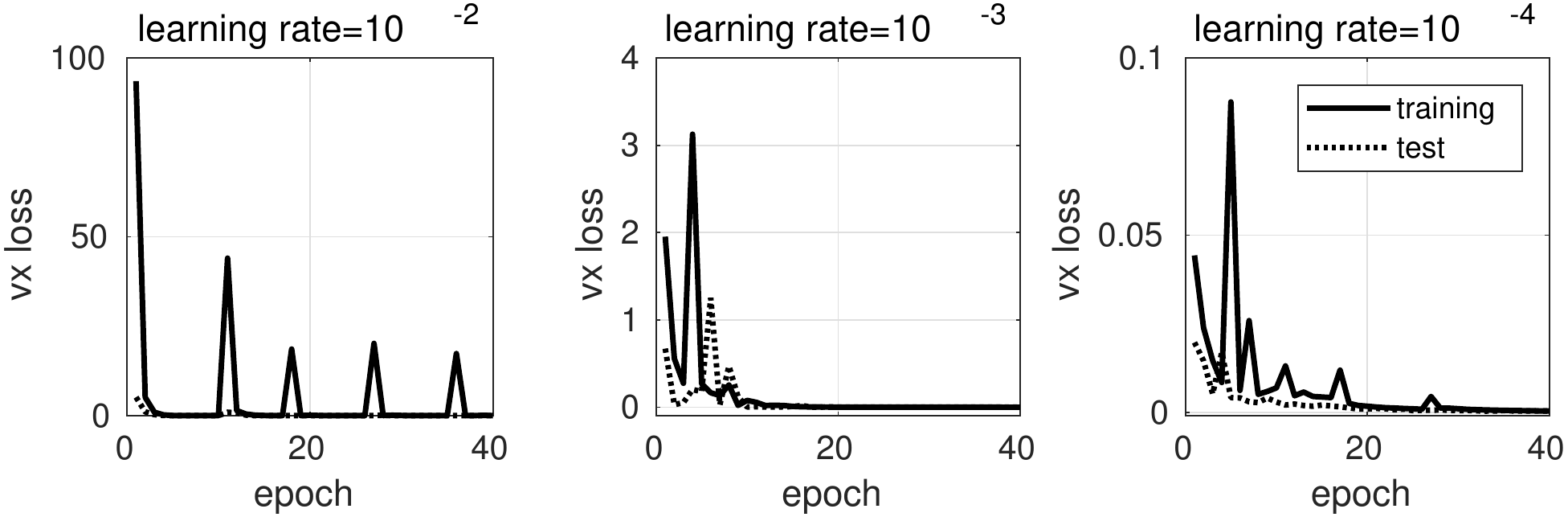}}
\caption{The learning curves of ARCH1 trained using different learning rates to extrapolate the 0-4Hz low frequencies of (a) $v_y$ and (b) $v_x$ from the 4-25Hz band-limited elastic recordings. }
\label{fig:benchmark_learningrate}
\end{figure}

The second essential hyperparameter is the learning rate of the optimizer. Figure~\ref{fig:benchmark_learningrate} compares the learning curves when the learning rates equal to $10^{-2}$, $10^{-3}$ and $10^{-4}$, respectively. If the learning rate is small, training is more reliable, but it will take significant time because steps towards the minimum of the loss function are tiny. The model may also miss the important patterns in the training data; Conversely, if the learning rate is high, training may not converge. Weight changes are so large that the optimizer overshoots the minimum and makes the loss worse. We observe that a learning rate of $10^{-3}$ seems to be an optimal learning rate and can quickly and stably find the minimum loss.   

\begin{figure}
\centering    
\subfigure[]{\label{fig:benchmark_others_vy_training_loss}\includegraphics[width=150mm]{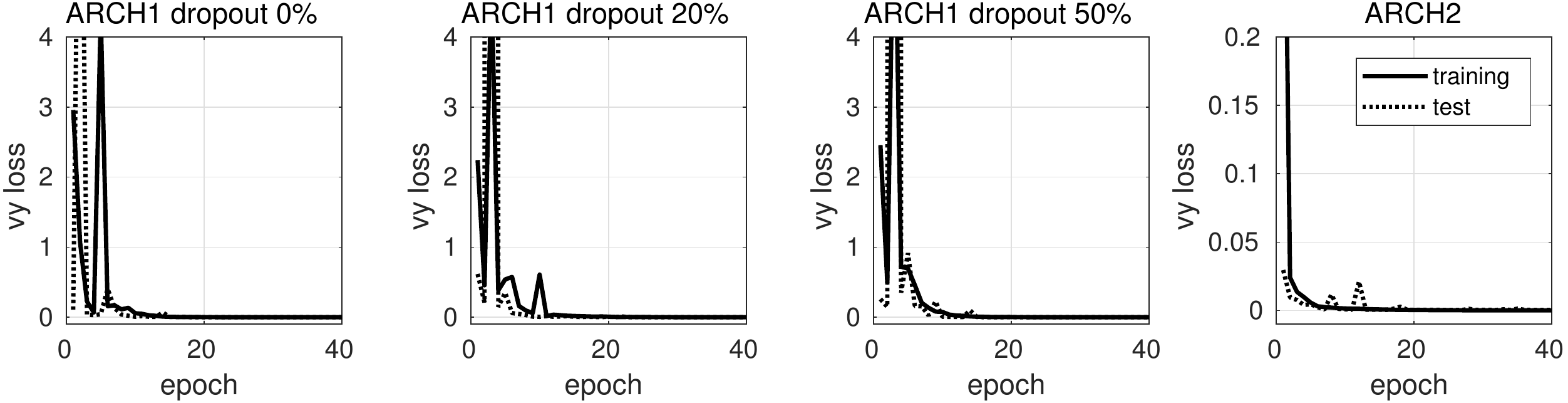}}
\subfigure[]{\label{fig:benchmark_others_vx_training_loss}\includegraphics[width=150mm]{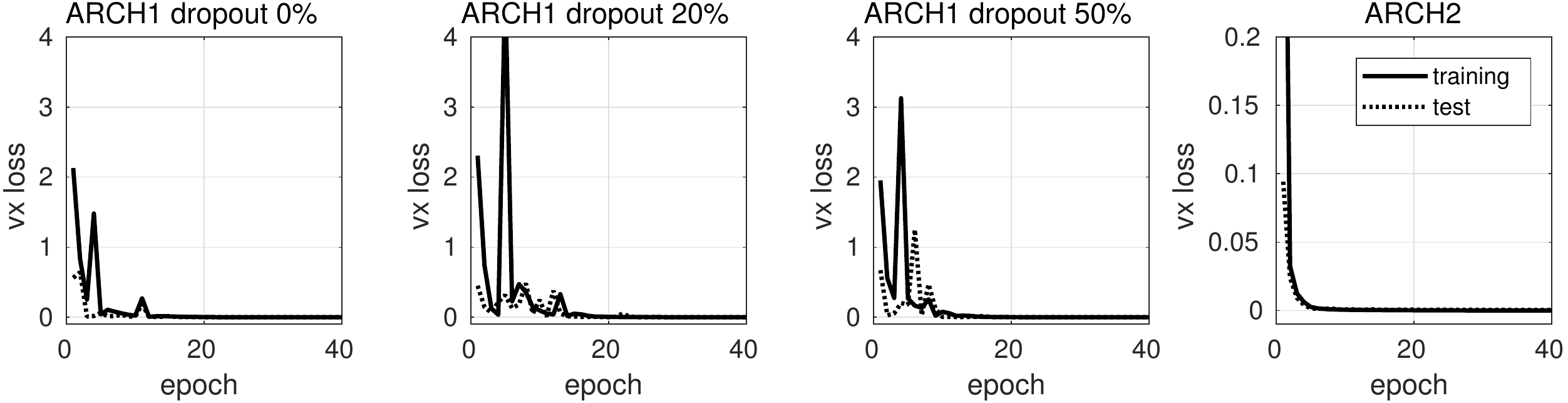}}
\caption{The learning curves of ARCH1 and ARCH2 trained to extrapolate the 0-4Hz low frequencies of (a) $v_y$ and (b) $v_x$ from the 4-25Hz band-limited elastic recordings.}
\label{fig:benchmark_others}
\end{figure}

Moreover, we study a model-related hyper-parameter, i.e., dropout in ARCH1. Dropout prevents neural networks from overfitting by randomly dropping neurons during training. Each neuron is retained with a fixed probability $p$ independent of other neurons. $p$ can be chosen using a validation set or can simply be set at 0.5, which seems to be close to optimal for a wide range of networks and tasks \citep{srivastava2014dropout}. Figure~\ref{fig:benchmark_others} compares the learning curves of ARCH1 with a dropout rate of $0\%$ (no dropout), $20\%$ and $50\%$, respectively. It seems that for this training dataset, a case without dropout layer does not hurt the performance of ARCH1. All of the three cases give reasonably right learning curves on both training and test datasets.

\begin{figure}
 \includegraphics[width=\columnwidth]{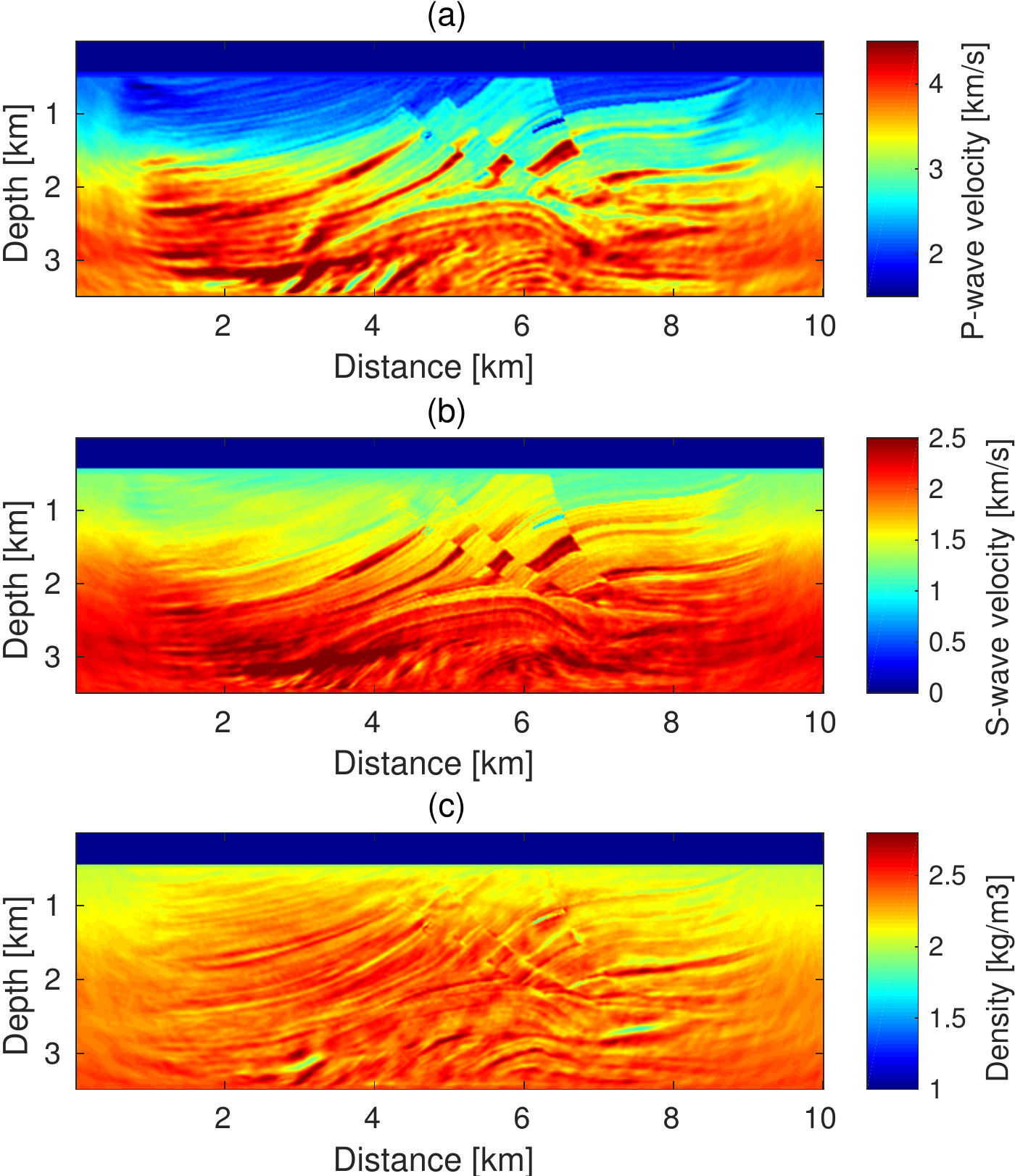}
   \caption{Comparison of the inverted (a) $\mathbf{v}_p$, (b) $\mathbf{v}_s$ and (c) $\rho$ models from extrapolated elastic FWI. The resulting models start from the inverted low-wavenumber models using $2.0-4.0Hz$ data extrapolated using ARCH2.}
   \label{fig:arch2_extra_inversion_result_stage_4_it_20}
\end{figure}

Finally, we compare the performance of ARCH1 and ARCH2 using the same training dataset and hyperparameter. Each network is trained twice with the training dataset of $v_x$ and the training dataset of $v_y$. According to the learning curves in Figure~\ref{fig:benchmark_others}, the training of ARCH2 is much more stable than that of ARCH1. ARCH2 also requires less training time, due to the less trainable parameters compared with ARCH1. Figure~\ref{fig:arch2_extra_inversion_result_stage_4_it_20} shows the extrapolated elastic FWI results started from the 2-4Hz extrapolated low frequency data using ARCH2. According to the comparison of the inverted models started from the extrapolated data using ARCH1 and ARCH2, both neural networks are able to provide sufficient accuracy for the inversion of $v_p$ and $v_s$.

\newpage
\section{Discussions and Limitations} 
\label{section:limitation}


Recovering the density using FWI is very challenging, independently of the bandwidth extension question, for the following reasons. (1) Cross-talk happens using short-offset data since P-wave velocity and density have the same radiation patterns at short apertures. (2) The variations in density are smaller than those in velocities. (3) Inversion of density requires ultra-low frequencies. Although elastic FWI does not always allow to correctly estimate density, it stands a better chance of properly reconstructing velocities, with either extrapolated or true low frequencies.


In extrapolated FWI, the choice of starting frequency is a trade-off between the accuracy of extrapolated low frequency data and the lowest frequency to mitigate the cycle-skipping problem. We start elastic FWI with 2-4Hz extrapolated low frequency data due to the insufficient extrapolation accuracy in the near-zero frequency range. The accuracy of the 2-4Hz extrapolated low frequency data is sufficient for elastic FWI of P-wave and S-wave velocities when starting from 4Hz band-limited data. However, the inversion of density model still surfers from the cycle-skipping problem and lack of the low-wavenumber structure.



The numerical example shows that the neural network cannot meaningfully generalize from the acoustic training dataset to the elastic test dataset. In addition to the wave propagation driven by different physics, another factor that makes the generalization fail could be numerical modeling. The synthetic elastic training and test datasets are simulated by solving the stress-velocity formulation of the wave equation using standard staggered grid with an eighth order FD operator. However, the acoustic training dataset is simulated by solving the stress-displacement formulation using a sixth order FD operator. 

The source signal is assumed to be known for extrapolated elastic FWI in this paper. However, for field data, the source signal may vary shot by shot. One solution could be to retrieve the source wavelet of the field dataset firstly, and then artificially boost the low-frequency energy after denoising. The new source signal can be used to synthesize the training dataset for low-frequency extrapolation. It can also be the source wavelet in the following elastic FWI using the extrapolated low-frequency data. In this way, the uncertainty of the source can be controlled to some extent.

We do not provide the numerical results of other factors that affect the performance of the deep learning models. For example, regularization of training loss, number of iterations, parameters of neural network, number of training samples and even inverse crime. Deep learning models contain many hyper-parameters and finding the best configuration for these parameters in a high dimensional space is challenging.

Finally, neural networks are trained using a stochastic learning algorithm. This means that the same model trained on the same dataset may result in a different performance. The specific results may vary, but the general trend should be the same, as reported in Section~\ref{section:benchmark}.

\section{Conclusions}
\label{section:conclusion}

To relieve the dependency of elastic FWI on starting models, low-frequency extrapolation of multi-component seismic recordings is implemented to computationally recover the missing low frequencies from band-limited elastic data. The deep learning model is designed with a large receptive field in two different ways. One directly uses a large filter on each convolutional layer, the other utilizes dilated convolution to increase the receptive field exponentially with depth. By training the neural network twice, once with a dataset of horizontal components and once with a dataset of vertical components, we can extrapolate the low frequencies of multi-component band-limited recordings separately. The extrapolated 0-5Hz low frequencies match well with the true low-frequency data on the Marmousi2 model. Elastic FWI using 2-4Hz extrapolated data shows similar results to the true low frequencies. The accuracy of the extrapolated low frequencies is enough to provide low-wavenumber starting models for elastic FWI of P-wave and S-wave velocities on data band-limited above 4Hz.

The generalization ability of the neural network over different physical models is studied in this paper. The neural network trained on purely acoustic data shows larger prediction error on elastic test dataset compared to the neural network trained on elastic data. Therefore, collecting more realistic elastic training dataset will help to process the field data with strong elastic effects.



\newpage

\end{document}